\documentclass{aa}  

\usepackage{graphicx}
\usepackage{txfonts}
\usepackage{lipsum}
\usepackage{subcaption}         
\usepackage{lscape}             
\usepackage{placeins}           

\usepackage{sidecap}

\def\targ{NGC7538~IRS1}

\def\wat{H$_{2}$O}
\def\meth{CH$_{3}$OH}
\def\nh3{NH$_{3}$}
\def\kms{km~s$^{-1}$}

\def\masyr{mas~yr$^{-1}$}

\def\Vlsr{$V_{\rm LSR}$}
\def\Jyb{Jy~beam$^{-1}$}

\def\G24{G24.78$+$0.08}
\def\HII{H{\sc ii}}

\newcommand{\ms}{$M_{\odot}$}
\newcommand{\ls}{$L_{\odot}$}
\newcommand{\msyr}{$M_{\odot}$~yr$^{-1}$}
\newcommand{\pas}{$\rlap{.}^{\prime\prime}$}
\newcommand{\pss}{$\rlap{.}^{\rm s}$}

\newcommand{\degree}{$^{\circ}$}

\begin{document}

\title{Protostellar Outflows at the EarliesT Stages (POETS). VII. Circumstellar gas kinematics traced by water masers inside the HC~\HII\ region \targ}
   
  \titlerunning{Gas kinematics inside HC~\HII\ regions via water masers }

   \author{L. Moscadelli \inst{1}
          \and
           C. Goddi \inst{2,3,4}
          \and
       T. Hirota \inst{5,6}
          \and
          A. Sanna \inst{4}  
          }

   \institute{INAF-Osservatorio Astrofisico di Arcetri, Largo E. Fermi 5, I-50125, Firenze, Italy \\  \email{luca.moscadelli@inaf.it}
         \and
      Dipartimento di Fisica, Universit\'a degli Studi di Cagliari, SP Monserrato-Sestu km 0.7, I-09042 Monserrato (CA), Italy  
         \and
  Instituto de Astronomia, Geofísica e Ciências Atmosféricas, Universidade de São Paulo, R. do Matão, 1226, São Paulo, SP 05508-090, Brazil       
         \and
   INAF - Osservatorio Astronomico di Cagliari, Via della Scienza 5, 09047 Selargius (CA), Italy
         \and
  National Astronomical Observatory of Japan, 2-12  Hoshigaoka, Mizusawa, Oshu, Iwate 023-0861, Japan
         \and
SOKENDAI (The Graduate University for Advanced Studies), 2-21-1 Osawa, Mitaka, Tokyo 181-8588, Japan
             }

 
  \abstract
   {The hyper-compact~(HC)~\HII\ region phase of a newly born massive star is presently poorly understood, particularly in relation to how the enhanced UV radiation impacts the kinematics of the surrounding gas and affects mass accretion.}
   {This article focuses on \targ, one of the most luminous and studied HC~\HII\ regions in the northern hemisphere. Our aim is to identify the young stellar objects (YSOs) embedded within the ionized gas and study their nearby kinematic structures. This work expands on a recent survey called Protostellar Outflows at the EarliesT Stages (POETS), which has been devoted to studying young outflow emission on scales of 10--100~au near luminous YSOs, before they start photoionizing the surrounding medium.} 
   {We carried out multi-epoch Very Long Baseline Array (VLBA) observations of the 22~GHz water masers toward \targ\ to measure the maser 3D velocities, which, following POETS' findings, are reliable tracers of the protostellar winds. Recently, we reobserved the water masers in \targ\ with sensitive global very long baseline interferometry (VLBI) observations to map weaker maser emission and also study the maser time variability.}
   {Our study confirms the presence of two embedded YSOs, IRS1a and IRS1b, at the center of the two linear distributions of 6.7~GHz methanol masers observed in the southern and northern cores of the HC~\HII\ region, which have been previously interpreted in terms of edge-on rotating disks. The water masers trace an extended ($\ge$~200~au) stationary shock front adjacent to the inner portion of the disk around IRS1a. This shock front corresponds to the edge of the southern tip of the ionized core and might be produced by the interaction of the disk wind ejected from IRS1a with the infalling envelope. The water masers closer to IRS1b follow the same local standard of rest (LSR) velocity (\Vlsr) pattern of the 6.7~GHz masers rotating in the disk, but the direction and amplitude of the water maser proper motions are inconsistent with rotation. We propose that these water masers are tracing a photo-evaporated disk wind, where the maser \Vlsr\ traces mainly the disk rotation and the proper motions the poloidal velocity of the wind. Finally, a sensitive  NSF’s Karl G. Jansky Very Large Array (JVLA) 1.3~cm image of the HC~\HII\ region obtained from archival data reveals a disk-jet system, illuminated by the UV radiation from IRS1a, associated with an YSO, IRS1c, placed $\approx$~0\pas5 (or $\approx$~1350~au) to the south of the ionized core.}
   {This work shows that VLBI observations of the 22~GHz water masers can be used to trace disk winds near ionizing YSOs embedded within compact \HII\ regions.}


   \maketitle

\section{Introduction}

\nolinenumbers

Observations and models indicate that the formation of stars within the mass range \ 0.1--30~\ms\  proceeds through a disk-outflow system in which mass accretion and ejection are intimately related. Nevertheless, a fundamental difference between low- ($\sim$1~\ms) and high-mass ($\gtrsim 8$~\ms) stars is that the latter start hydrogen nuclear burning while still accreting gas from the surrounding disk and envelope. The enhanced UV radiation of a newly born massive star ionizes and heats the circumstellar gas, which could in principle result in a rapid expansion of an ionized bubble and the formation of an \HII\ region. This classical Str\"omgren expansion model is challenged by the large number of hyper- (HC; size $\lesssim$ 0.01~pc) or ultra-compact (UC;  $\lesssim$  0.1~pc) \HII\ regions in the Galaxy \citep[e.g.,][]{Woo89}, which indicates that the duration of the compact phase of the \HII\ region should last about ten times longer than expected. Considering that massive young stellar objects (YSOs) are characterized by high accretion rates of $\gtrsim 10^{-4}$~\msyr\ \citep[e.g.,][]{Tan14}, \citet{Ket03} suggested that mass infall could confine the \HII\ region inside a small radius until the stellar mass grows large enough to boost the Lyman continuum and cause a rapid increase in the radius of the ionized gas. The stellar mass for the onset of an expanding \HII\ region would depend  on  the  geometry  and  mass  accretion  rate  onto  the (proto)star  \citep{Hos10},  and  it is predicted to vary in the range of\ 10--30~\ms.

The observation of hydrogen radio recombination lines (RRLs) allows us to investigate gas motions and physical conditions inside HC- or UC~\HII\ regions. The required high angular resolution of \ $\le$~0\pas1 and good sensitivity can be achieved both at centimeter  and millimeter wavelengths, using the  NSF’s Karl G. Jansky Very Large Array (JVLA) and the Atacama Large Millimeter/submillimeter Array (ALMA), respectively, covering an extended range  of RRL quantum numbers ($\approx$~30--110). In several objects, these observations have discovered large ordered motions across the ionized gas on scales of 100--1000~au, corresponding to infall toward, rotation around, or outflow away from the ionizing star(s) \citep{Sew08,Ket08b,DeP20,Riv20,ZhaY19}. These global motions of the ionized gas suggest that the combination of mass accretion and ejection onto and away from the star can also proceed after the onset of photoionization, in agreement with the  theoretical work by \citet{Ket03} predicting the existence of trapped HII regions. In this regard, a direct example is that of the HC~\HII\ region in G24.78$+$0.78 \citep{Mos21}, where an ionized disk-outflow system is surrounded by an outer molecular disk still actively accreting from the hosting molecular core.

   \begin{figure}
   \centering
   \vspace*{-1cm}\includegraphics[width=0.5\textwidth]{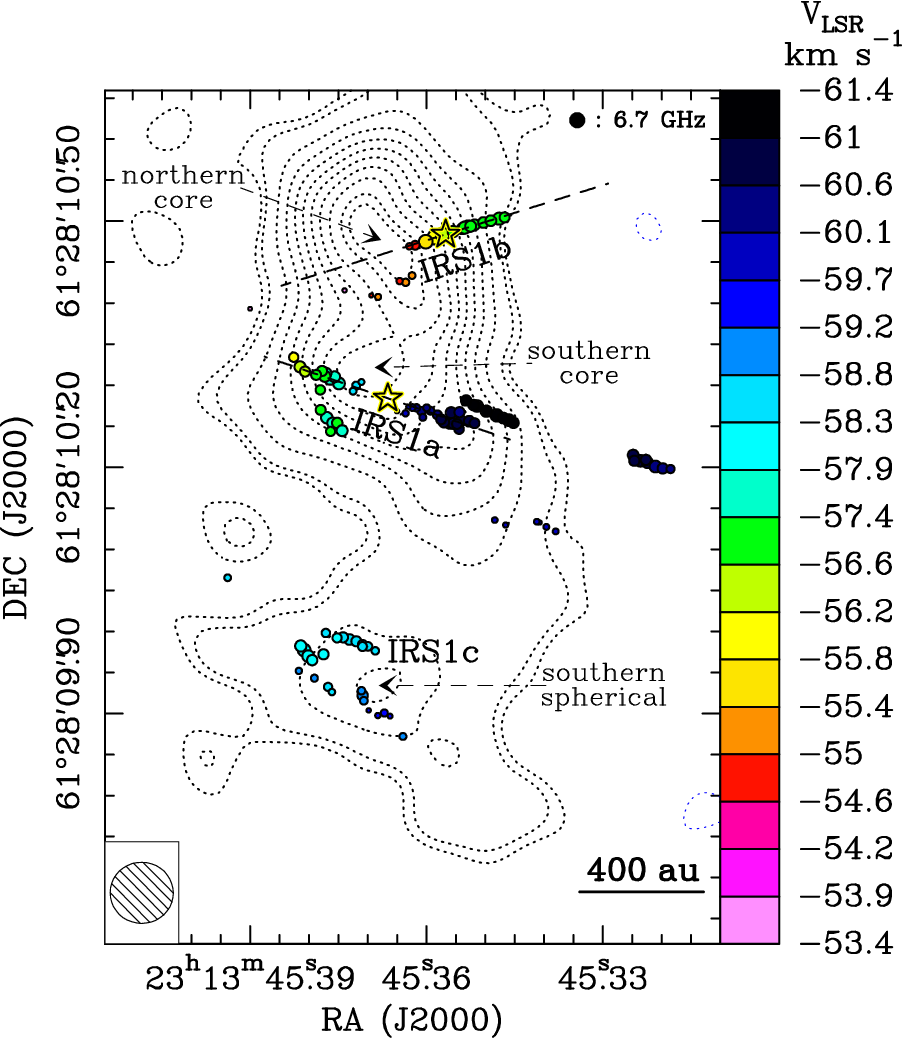}
      \caption{6.7~GHz \meth\ masers detected over three epochs with the European VLBI Network (EVN), overlaid  on the uniformly-weighted image of the 1.3~cm continuum observed with the VLA A-Array in December 1992 \citep{Mos14}.
Colored dots show the absolute position  
of individual maser features,
with colors denoting the maser \Vlsr\ 
according to the color-velocity conversion code
reported on the right side of the panel. The dot area is proportional to the logarithm of the maser
intensity.
The 1.3~cm map (dotted contours) was produced using archival data (VLA exp. code: AG360) originally reported by \citet{Gau95}.
Plotted contours are \ 7\%, 10\% to 90\% (in steps of 10\%), and 95\% of the map
peak equal to 0.022~\Jyb. The 1$\sigma$ rms noise is $\approx$0.4~mJy~beam$^{-1}$. Dashed arrows point to the main 1.3~cm continuum peaks,
which are named following the notation by \citet{Gau95}.
The beam of the VLA-A 1.3~cm observations is reported in the inset
in the bottom left of the panel. Absolute positions are relative to the epoch March 2, 2017, which is the first of the four VLBA epochs of the BM454 22~GHz maser observations (see Sect.~\ref{vlba}).
The two dashed lines
indicate the major axes of the elongated distributions of 6.7~GHz masers in the northern and southern cores, and the two black-yellow stars labeled IRS1a and IRS1b mark the YSO positions determined by modeling the maser kinematics with edge-on disks in centrifugal equilibrium \citep{Mos14}. After \citet{Mos14}, a third YSO, labeled IRS1c, is responsible for the excitation of the continuum and 6.7~GHz maser emission in the southern spherical region.}
         \label{MG}
   \end{figure}
%

This article focuses on the HC~\HII\ region \targ, one of the most luminous -- $\sim$~10$^5$~\ls\ \citep{Aka05}, at a distance of 2.7~kpc \citep{Mos09} -- and studied high-mass star forming regions in the northern hemisphere.
The structure of the HC~\HII\ region (see Fig.~\ref{MG}) consists of two intense radio lobes separated by 0\pas2 (the "northern" and "southern" cores) and a weaker spherical component ("southern spherical") located $\approx$~0\pas5 to the south of the double core \citep{Cam84,Gau95}.  Hydrogen RRLs show extremely broad line widths ($\approx 250$~\kms) and multiple emission peaks varying with position across the core region, suggestive of different clumps of ionized gas along the line of sight moving at large relative velocities \citep{Gau95,Ket08}. Interferometric observations at millimeter wavelengths with angular resolutions ranging from  0\pas2 to 3$^{\prime\prime}$ \ have detected several hot-core molecules showing inverse P-Cygni spectral profiles, probing inward motions at scales of \ $\gtrsim 1000$~au with a mass infall rate of \ $\sim$~10$^{-3}$~\msyr\ \citep{Qiu11,Zhu13,Beu13}. At scales of a few 10$^{\prime\prime}$, a molecular outflow is observed emerging from \targ\ in the southeast-northwest direction \citep{Kam89,Dav98,Qiu11}, which has a similar orientation to the elongated mid-infrared emission threading the HC~\HII\ region at scales of $\sim$~1$^{\prime\prime}$ \citep{DeBui05}.

The double-lobe morphology of the core has originally been interpreted in terms of an ionized jet ejected in the north-south direction by a single O star at the center of the core \citep{Cam84,Gau95}, which is accreting mass at a rate high enough to quench the HC~\HII\ region. Subsequently, the observation of two linear distributions of intense 6.7~and~12~GHz CH$_3$OH (methanol) masers, near the northern and southern cores, presenting a regular change in the local standard of rest (LSR) velocity (\Vlsr) with position (see Fig.~\ref{MG}), has led to an alternative interpretation postulating the presence of two distinct YSOs surrounded by (almost) edge-on disks\footnote{In this article, the term “disk” is used to generally mean circumstellar gas close to the equatorial plane rotating in centrifugal equilibrium, independently on the range of rotation radii and the profile of mass distribution.} traced by the methanol maser emission \citep{Mos14}. More recently, JVLA observations of thermal methanol emission at a linear resolution of $\approx$~150~au have revealed two velocity gradients across the ionized core of the HC~\HII\ region whose orientation, position, and extent agree well with the ones traced with the 6.7~GHz methanol masers at linear resolutions of 1~au, supporting the existence of two embedded disk-like structures \citep{Beu17}.
 Very recently, on the basis of VLA and Combined Array for Research in Millimeter-wave Astronomy (CARMA) observations, covering an extended wavelength range between 1.3~and~7~mm and achieving a maximum linear resolution of \ $\approx$~300~au, \citet{Sand20} have argued against the multiple-disk interpretation, confirming that \targ\ is ionized by a single O star.

To further investigate the nature of \targ\ and resolve the gas kinematics of the YSO(s) embedded in the HC~\HII\ region, in this work we employ sensitive multi-epoch very long baseline interferometry (VLBI) observations of the 22~GHz water masers. Our previous studies \citep{Mos07,San10b,Mos11a,God11a} have demonstrated the ability of 22~GHz maser VLBI to trace the velocity structure of the outflows near YSOs, and the ``Protostellar Outflows at the Earliest Stages'' (POETS) survey \citep{Mos16,San18,Mos19} has recently imaged through water masers the innermost outflow ejection on scales of \ 10--100~au in a statistically significant sample (37) of luminous YSOs.
Among the main results of POETS, we find that the water masers are always associated with weak, slightly resolved continuum emission, tracing the thermal jet emitted from the YSO, and the 3D velocity distribution of the water masers in most of the sources of the sample can be interpreted in terms of a magnetohydrodynamic disk wind. In POETS, we have targeted the earliest phases of star formation, when the YSO has not grown enough in mass to photoionize the surrounding gas. This article expands on the POETS study by targeting  an HC~\HII\ region. We have used multi-epoch Very Long Baseline Array (VLBA) observations to measure the 3D velocities of the water masers brighter than 100~m\Jyb\ and more sensitive global VLBI observations to map the emission of the weakest masers. In Sect.~\ref{vlbiwat}, we describe the VLBI observations, and the data calibration and analysis is reported in Sect.~\ref{calana}. Sect.~\ref{res} presents the distribution of the water maser positions and 3D velocities. The discussion of the new results is provided in Sect.~\ref{dis} and the conclusions are drawn in Sect.~\ref{conclu}.

\section{VLBI water maser observations} 
\label{vlbiwat}

We observed the  $6_{16} - 5_{23}$ H$_2$O maser transition \citep[rest frequency 22.2350798 GHz,][]{Pick98} toward \targ\ with both the VLBA and the global VLBI array. To determine the absolute positions and velocities, we performed phase-reference observations, alternating scans on the target and the phase-reference calibrators every 30--45~s; the phase-reference calibrators were the quasars \ J2254$+$6209 (separation from the maser target of 2.38\degree, 8~GHz correlated-flux of \ $\approx$30~mJy), J2302$+$6405 (2.92\degree,  $\approx$100~mJy) and J2339$+$6010 (3.38\degree,  $\approx$180~mJy). Calibration scans of 3--10~min were observed every few hours on the fringe-finder and bandpass calibrators \ J2202$+$4216, J2005$+$7752,  J1638$+$5720, J2236$+$2828, J0102$+$5824, and J0555$+$3948.

We recorded dual circular polarization through four adjacent bandwidths of 16~MHz, one of them centered at the target systemic \Vlsr \ of \ $-57.0$~\kms. The four 16~MHz bandwidths were used to increase the signal-to-noise ratio (S/N) of the weak continuum (phase-reference) calibrators. The VLBA  data were correlated with the VLBA-DiFX software correlator in Socorro (New Mexico, USA), and the global VLBI data with the SFXC correlator at the Joint Institute for VLBI in Europe (JIVE, in Dwingeloo, the Netherlands). For each observation, we employed two correlation passes: \ 1)~2000 (for VLBA) and 1024 (for global VLBI) spectral channels to correlate the maser 16~MHz bandwidth; and \ 2)~32 spectral channels to correlate the whole set of four 16~MHz bandwidths. The spectral resolution attained across the maser 16~MHz band was 0.11~and~0.21~\kms\ for the VLBA and global VLBI observations, respectively. The correlators' averaging time was 1~s. In the following, we report the details of both observations separately.

\subsection{VLBA observations}
\label{vlba}

We observed \targ\ 
(tracking center: RA(J2000) = $23^{\rm h} \, 13^{\rm m}$ 45\pss3626, Dec(J2000) = $+61$\degree\ $28^{\prime}$ 10\pas508) with the ten antennae of the  VLBA (exp. code: BM454) of the National Radio Astronomy Observatory (NRAO\footnote{NRAO is a facility of the National Science Foundation operated under cooperative agreement by Associated Universities, Inc.}) at four epochs: March 2, May 27, June 11, and July 22, 2017; each epoch lasted 14~hr. Because of technical problems, the antennae of Mauna Kea and Pie Town did not take part in the observations at the first epoch. We placed  “geodetic” blocks \citep{Rei09} before the start, in the middle, and after the end of the phase-reference observations, in order to model and remove uncompensated interferometric delays introduced by the Earth's atmosphere and improve the astrometric accuracy. Due to the different antennae available at each epoch, the full width at half maximum (FWHM) major and minor sizes of the beam, using natural weighting, vary over the observing epochs in the ranges\ 0.6--1.2~mas and 0.3--0.8~mas, respectively, and the beam position angle (PA) in the range \  [$-66$\degree, 75\degree]. In channel maps with a (relatively) weak signal, the 1$\sigma$ root mean square (rms) noise varies within \ 7--11~mJy~beam$^{-1}$, which is close to the expected thermal noise of $\approx$~7~mJy~beam$^{-1}$.

\subsection{Global VLBI observations}
\label{glob}

We observed \targ\ 
(tracking center: RA(J2000) = $23^{\rm
h}  \, 13^{\rm m}$ 45\pss37,  Dec(J2000) = $+61$\degree\ $28^{\prime}$ 10\pas35) with global VLBI for 24~hr, starting on June 5, 2023, at 19:00~UT. A total of 17 antennae were involved in the observations and provided useful data: eight antennae of the European VLBI network (EVN\footnote{The European VLBI Network is a joint facility of independent European, African, Asian, and North American radio astronomy institutes.  Scientific results from data presented in this publication are derived from the following EVN project code: GM082.}),
Medicina, Jodrell\_Bank, Effelsberg, Onsala, KVN\_Ulsan, KVN\_Yonsei, Urumqi, and Tianma; plus nine antennae of the VLBA, Brewster, Fort Davis, Hancock, Kitt Peak, Los Alamos, Mauna Kea, North Liberty, Pie Town, and Saint Croix. While the EVN antennae observed the target only, the VLBA also performed phase-reference observations (over 10~hr). During the phase-reference session, the target and the calibrators were always observed by the VLBA simultaneously with the EVN to ensure global VLBI baselines.  Using natural weighting, the FWHM major and minor sizes of the beam are \ 0.24~mas and 0.20~mas, respectively, and the beam PA is \  $-10.7$\degree. In channel maps with a (relatively) weak signal, the 1$\sigma$ rms noise is $\approx$1~mJy~beam$^{-1}$, equal to the expected thermal noise.

\section{Data calibration and analysis}
\label{calana}

\subsection{VLBI}
\label{VLBI}

Data were reduced with the Astronomical Image Processing System \citep[\textsc{AIPS},][]{Gre03} package following the VLBI spectral line procedures in the \textsc{AIPS} COOKBOOK\footnote{\url{http://www.aips.nrao.edu/cook.html}}. At each VLBI epoch, the emission of an intense and compact maser channel was self-calibrated, and the derived (amplitude and phase) corrections were applied to all maser channels before imaging. Water maser emission has been searched for over images extending  \ 1\pas1 \ in both \ RA $\cos \delta$ \ and \ DEC, and 84~\kms \ in \Vlsr.
For a description of the criteria used to identify individual maser features, derive their parameters (position, intensity, flux, and size), and measure their absolute proper motions, we refer to \citet{Mos06}. Individual maser features are a collection of quasi-compact spots observed on contiguous channel maps and spatially overlapping (within their FWHM size). The positions of different spots are determined by fitting a two-dimensional elliptical Gaussian to their brightness distribution. For a detailed discussion of the method of calculating the error in the relative positions of spots and features, we refer to \citet{Mos24}.

Tables~\ref{wat1}~and~\ref{wat2} report the parameters (intensity, \Vlsr, position, and absolute proper motion) of the 22~GHz water masers in \targ\ from the VLBA and global VLBI observations, respectively.
For both observations and at each epoch, inverse phase referencing \citep{Rei09} produced good (S/N $\ge$~10) images of the phase-reference calibrators. Taking into account that the calibrators are relatively compact, with sizes of $\lesssim$~1~mas, and that the absolute position of the calibrators is known within a few 0.1~mas, we estimate that the error on the absolute position of the masers is \ $\lesssim$~0.5~mas. 
Correcting for the uncompensated delay introduced by the Earth's atmosphere \citep{Rei09}, our VLBA observations reach an astrometric accuracy in the maser-referenced calibrator images of \ $\lesssim$~0.05~mas, and the absolute proper motions have typical errors of \ 4~\kms.  
The derived absolute proper motions were corrected for the apparent motion due to the combination of the Earth's orbit around the Sun (parallax), the Solar Motion and the differential Galactic Rotation between our LSR and that of the maser source. The Solar Motion and the differential Galactic Rotation are well approximated by the proper motion of the 12~GHz methanol masers in \targ\ \citep[$-$2.45~\masyr\ in RA and $-$2.44~\masyr\ in DEC,][]{Mos09}, which approximately comove \citep[within 5~\kms,][]{Mos14} with the YSO(s) inside the \HII\ region.

\subsection{JVLA archival data}
\label{arch}

\citet{Beu17} used the JVLA in the most extended A-configuration to observe the 1.3~cm continuum emission toward \targ\ on 
June 21, 2015 (JVLA proposal ID: 15A-115). We downloaded this dataset from the NRAO archive and newly calibrated the observations by running the Common Astronomy Software Applications (\textsc{CASA}) Very Large Array (VLA) pipeline (version 6.5.4). To map the structure of the compact ($\lesssim $~0\pas3) ionized core, we produced a uniformly weighted image of the 1.3~cm continuum whose beam has FWHM major and minor sizes of \ 61~mas and 47~mas, respectively, and a PA at \  $-23.2$\degree.  To map the emission of the HC~\HII\ region extending up to $\approx$1$^{\prime\prime}$, we produced an image using Briggs weighting (robust 0), obtaining a beam with FWHM major and minor sizes of \ 80~mas and 62~mas, respectively, and a PA at \  $-23.3$\degree. Thanks to the large recording bandwidth of $\approx$2~GHz, the rms noise of $\sim$~10~$\mu$\Jyb\ of these images is about one order of magnitude smaller than that of the corresponding images from the 1992 VLA observations by \citet{Gau95}, which recorded a bandwidth of 25~MHz. As was reported by \citet{Beu17}, the absolute position of the phase calibrator employed in the 2015 JVLA observations was incorrect. Since the relative distribution of the 6.7~GHz masers has not changed over $\approx$30~yr, we can reasonably assume that the maser positions have also not changed in time with respect to the radio continuum peaks. Under this assumption, we have applied an offset of \ $\Delta$RA = $-$75~mas and $\Delta$DEC = $+$215~mas to align the “southern spherical” emission of our 1.3~cm continuum images with the 6.7~GHz maser cluster placed $\approx$0\pas5 to the south of the double-lobe core, in agreement with the 1992 VLA images (see Fig.~\ref{MG}).

\section{Results}

\label{res}

\begin{figure*}
   \centering
   \includegraphics[angle=0,width=\textwidth]{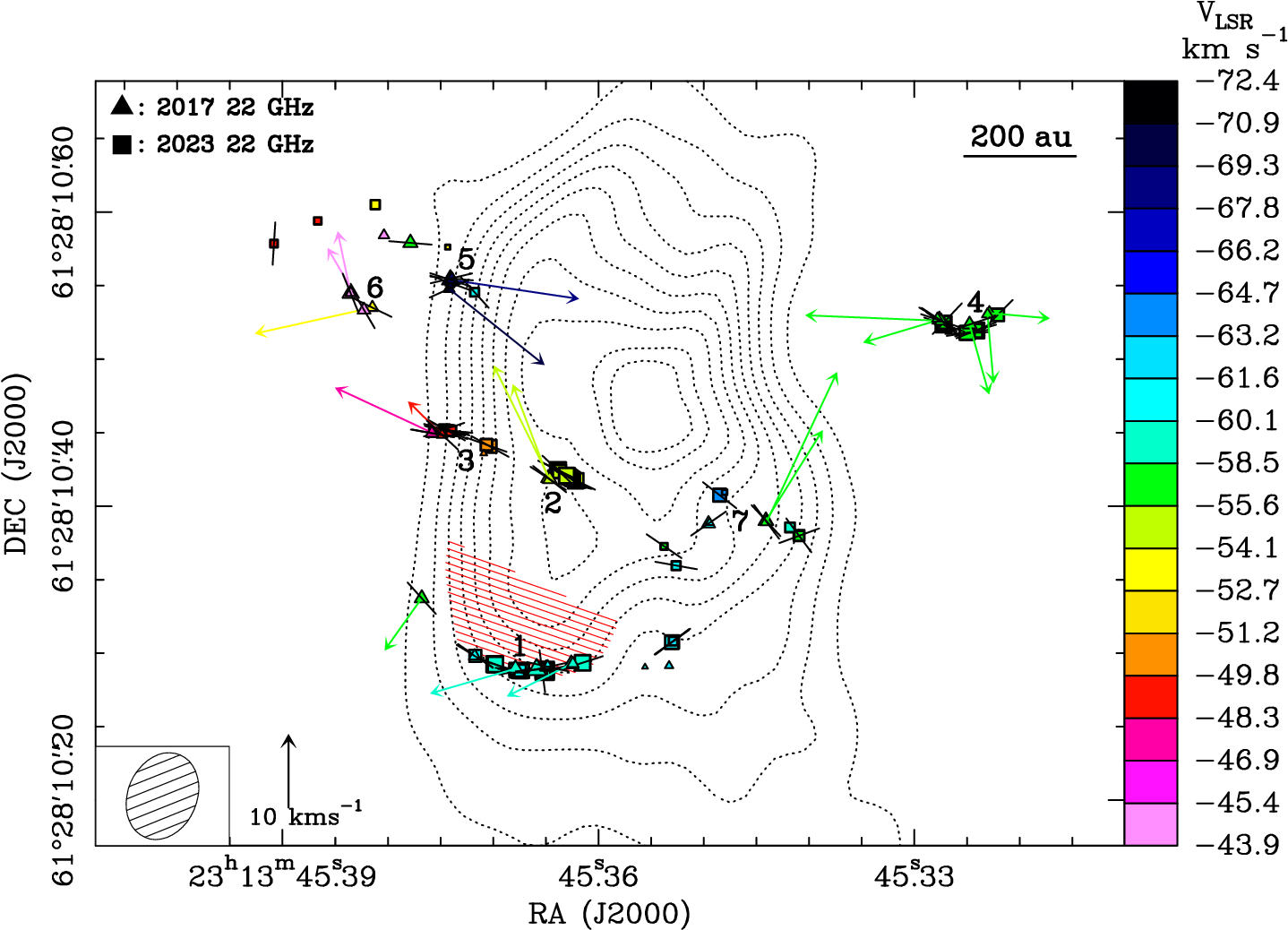}
      \caption{\wat\ masers (VLBA and global array) overlaid on the uniformly weighted image of the 1.3~cm continuum observed with the JVLA A-Array in June 2015. The absolute positions (relative to the epoch March 2, 2017) are reported with colored triangles and squares for the water masers from VLBA and global array observations, respectively. Colors denote the maser \Vlsr, and the symbol area is proportional to the logarithm of the maser
intensity. Colored arrows represent the absolute proper motions of the water masers measured through the four VLBA epochs, with the velocity scale shown in the bottom left of the panel. Only the proper motions with more accurate directions are plotted. Black segments overlaid on the water masers indicate the PA of the maser features. The main water maser clusters are labeled from 1 to 7. The 1.3~cm map (dotted contours) was produced using archival data  
(JVLA proposal ID: 15A-115, see Sect.~\ref{arch}) originally reported by \citet{Beu17}.
Plotted contours are \ 10\% to 90\% (in steps of 10\%), and 95\% of the map peak equal to 0.015~\Jyb. The southern tip of the ionized core, edged by the arc-like maser cluster~1, is hatched in red. The beam of the 1.3~cm image is reported in the inset in the bottom left of the panel.}
         \label{MBG}
   \end{figure*}

\begin{figure*}
   \centering
     \includegraphics[angle=0,width=\textwidth]{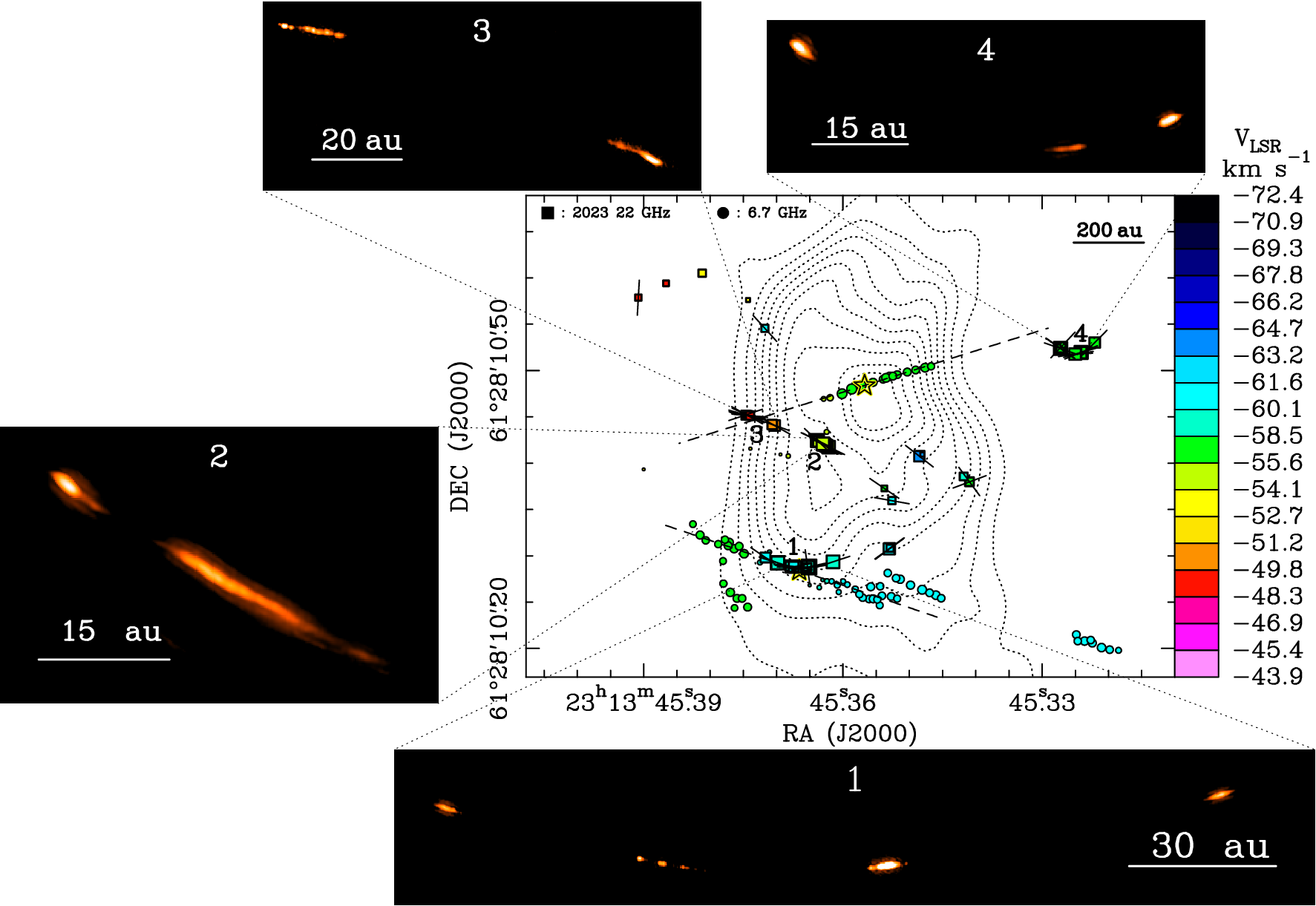}
      \caption{Structure of the water maser emission. Colored dots and the two dashed lines and black-yellow stars have the same meaning as in Fig.~\ref{MG}. Colored squares, black segments overlaid on the squares, and dotted contours have the same meaning as in Fig.~\ref{MBG}. The emission of the water maser clusters~1--4 is shown in the corresponding insets. To catch weaker maser features, plots of the maximum emission over the cluster velocity range are presented. The beam of the global VLBI observations has a FWHM size of 0.6~au and it is too small to be shown in the insets.}
         \label{G_stru}
   \end{figure*}

Fig.~\ref{MBG} shows the distribution and proper motions of the 22~GHz water masers in \targ\ overlaid on the 1.3~cm continuum emission from the reprocessed 2015 JVLA observations. In Fig.~\ref{G_stru}, we also plot the VLBI positions of the 6.7~GHz methanol masers by \citet{Mos14}. 
We have corrected maser and radio continuum positions for the apparent motion between the different observing epochs. This correction ensures that the relative positions among the different maser observations are accurate within 5~mas. The water masers detected in the four VLBA and single global VLBI epochs are concentrated in the same small (size $\approx$0\pas5) area close to the two linear distributions of 6.7~GHz masers that thread the core of the HC~\HII\ region (see Figs.~\ref{MBG}~and~\ref{G_stru}). The water masers can be grouped in seven main clusters (labeled in Fig.~\ref{MBG}) and, interestingly, the VLBA and global VLBI maser detections spatially overlap in clusters 1--4. 

 The shape of the maser emission can shed light on its birthplace. To this purpose, we derived the PA of the maser features using the expression:
\begin{equation}
 {\rm PA_f} = \frac{\sum_{i} \rm{PA}_i \; (\rm{Maj}_i \, / \, \rm{Min}_i)} { \sum_{i} \, (\rm{Maj}_i \, / \, \rm{Min}_i)} 
,\end{equation}
 
\noindent where the index, $i$, runs on the spots belonging to the feature, and \ $\rm{Maj}_i$, $\rm{Min}_i$, and PA$_i$ \ are the FWHM sizes along the major and minor axes, and PA of the spots, respectively (from the Gaussian fits, see Sect.~\ref{VLBI}). Figure~\ref{MBG} shows that the PAs of nearby features are similar and that the feature PA varies regularly with position. In particular, in clusters~1~and~4, which have an arc-like shape, the features are oriented parallel to the arc. In clusters~2~and~3, the maser brightness locally is elongated parallel to the overall cluster distribution. The maser emission of these four clusters is plotted in Fig.~\ref{G_stru}: it is evident that in each position the water maser emission follows a well-defined structure. 

Most of the water masers move at a speed of $\le$~15~\kms\ (see Fig.~\ref{MBG}), smaller than the typical values of 10--40~\kms\ measured for water masers associated with high-mass YSOs \citep{Mos20}. In the arc-shaped cluster~4, the water maser proper motions are directed approximately perpendicular to the arc (see Fig.~\ref{MBG}). In the other clusters, the masers with similar \Vlsr\ move in similar directions, although the direction of motion of the clusters changes significantly with the position. 

\begin{figure}
   \sidecaption
   \includegraphics[width=0.5\textwidth]{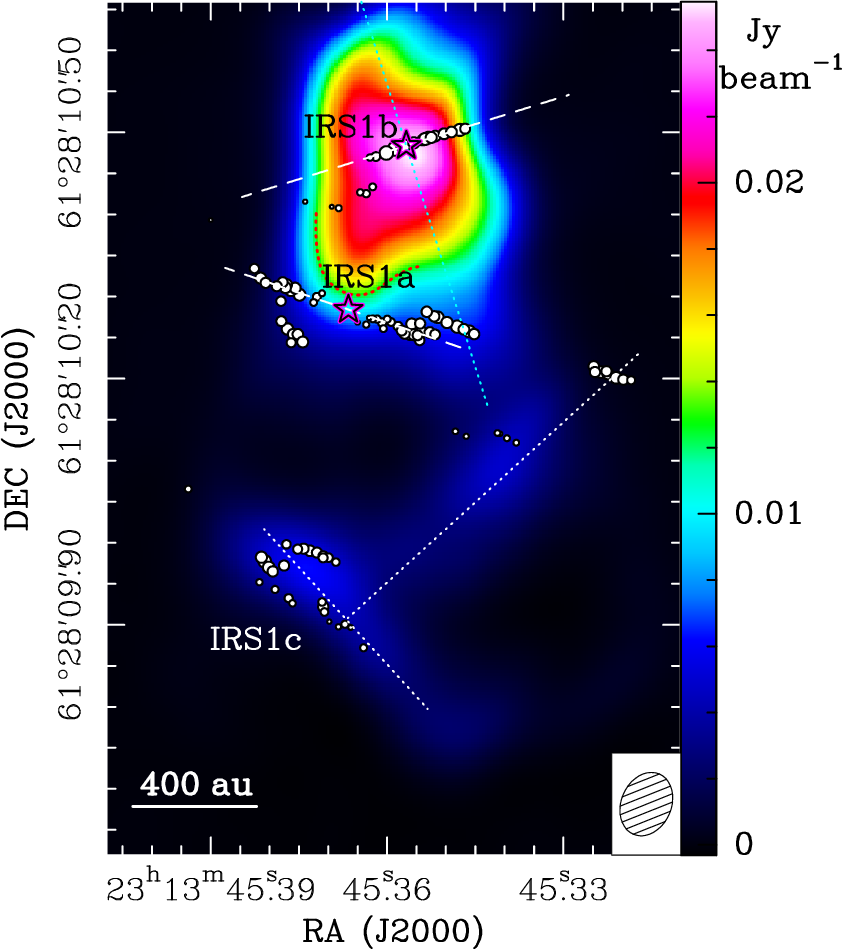}
      \caption{Structure of the HC~\HII\ region. The color map shows the image of the 1.3~cm continuum observed with the JVLA A-Array in June 2015, produced with Briggs weighting (robust 0). The beam of the 1.3~cm image is reported in the inset in the bottom right of the panel. The edge of the southern tip of the ionized core is marked with a dotted red curve. The dots, the two dashed lines, and the labeled stars have the same meaning as in Fig.~\ref{MG}. The dotted cyan and white lines draw the approximate axes of three elongated emissions discussed in the text.
      }
         \label{1_LS}
   \end{figure}

Comparing the 1992 VLA (Fig.~\ref{MG}) and 2015 JVLA (Figs.~\ref{MBG}~and~\ref{1_LS}) images of the HC~\HII\ region, the one-order-of-magnitude improvement in sensitivity results in three main morphological changes: \ 1)~the northern core of the VLA image shrinks into a more compact peak in the JVLA image; \ 2)~the edge of the southern core toward the south and southeast is better defined; and\ 3)~the extended southern spherical component of the VLA image is resolved by the JVLA observations in three elongated narrow emissions.

%

%

%

\section{Discussion}

\label{dis}

\subsection{Stationary shock fronts}
\label{inf}

\begin{figure*}
    \includegraphics[width=0.56\textwidth]{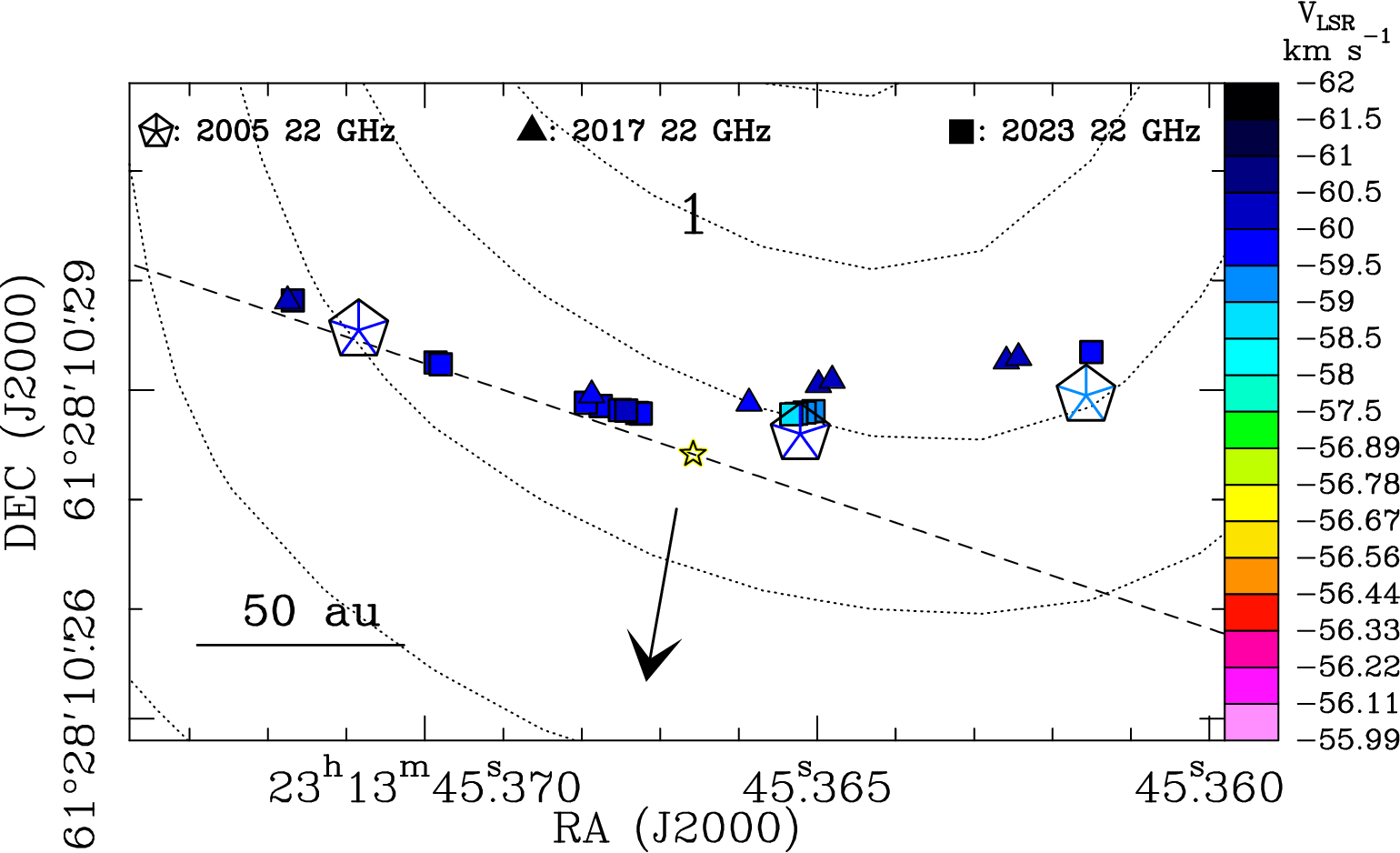} 
    \hspace*{0.3cm}\includegraphics[width=0.47\textwidth]{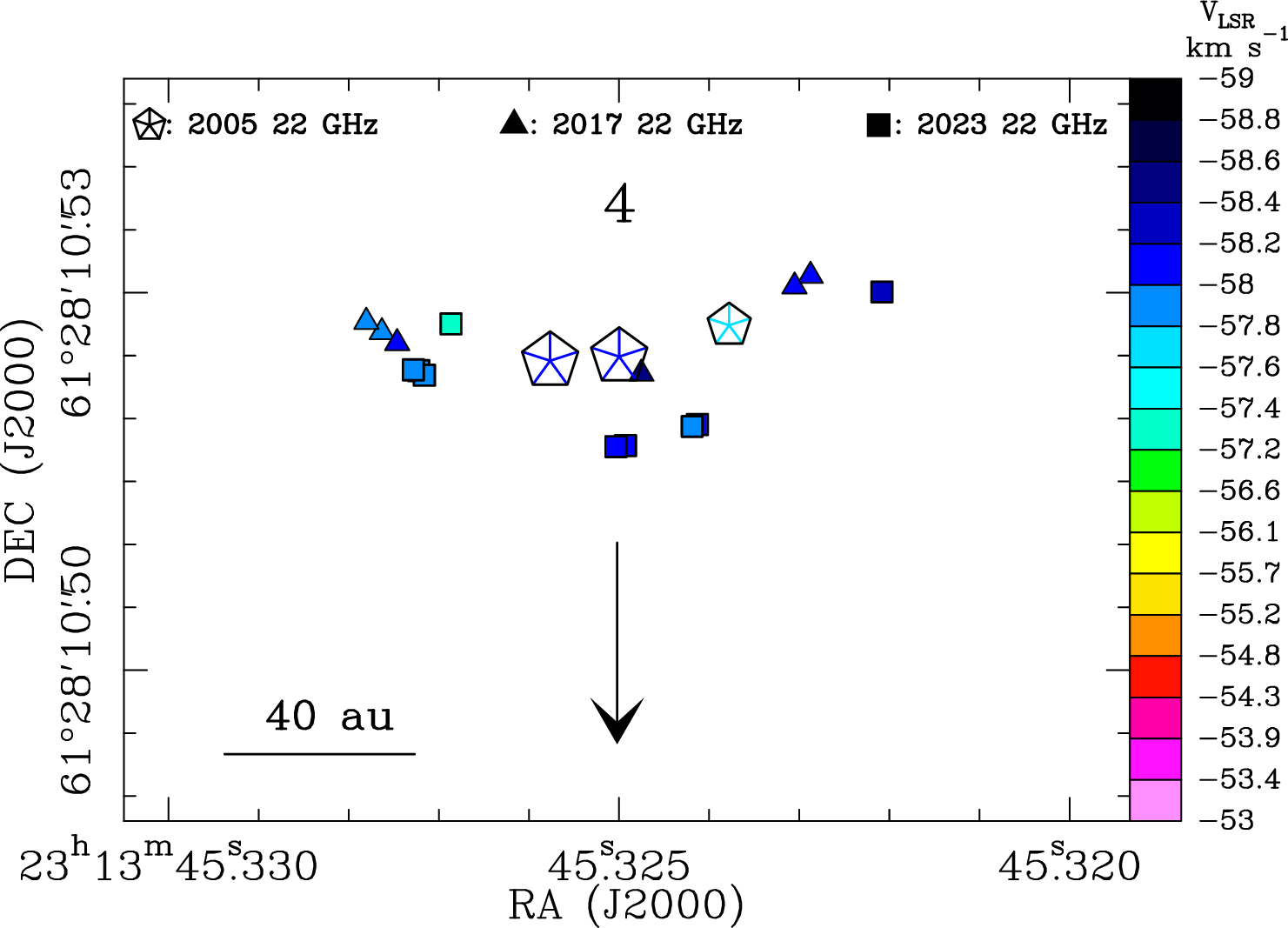}
      \caption{Comparison of water maser observations at different epochs. The left and right panels present an expanded view of the water maser clusters 1 and 4, respectively. The colored triangles and squares have the same meaning as in Fig.~\ref{MBG}. We use the same symbol size for all the masers to better emphasize their relative spatial distribution. The black-edged pentagons report the position of the water masers observed with EVN in November 2005 by \citet{Sur11b}. The big black arrows indicate the distance traveled by moving at a velocity of 10~\kms\ over 20~yr. In the left panel,  dotted contours and the dashed-line and black-yellow star have the same meaning as in Fig.~\ref{G_stru}.}
         \label{WM1-4}
   \end{figure*}

Considering the origin of the 22~GHz water masers in shocks \citep{Hol13}, it is reasonable to think that the water masers observed inside the HC~\HII\ region \targ\ arise in molecular material compressed and heated by its relative fast motion with respect to the ionized gas. The spatial correspondence in clusters~1-4 of the water masers observed in 2017 (VLBA) and 2023 (global VLBI) indicates the persistence of the conditions for maser excitation at those positions (see Fig.~\ref{MBG}). In particular, the arc-like shape of clusters~1~and~4 suggests that the masers trace extended ($\ge$~100~au) shock fronts at the interface between the ionized gas and molecular material. The lack of ionized gas at the position of cluster~4 could just be apparent because a more extended, low-level plateau of continuum emission could be filtered out by the long baselines of the JVLA in A-configuration. \citet{Sur11b} observed the 22~GHz water masers in \targ\ with the EVN in 2005 and most of their detections are found close to clusters~1~and~4, despite the uncertainty of $\approx$50~mas in the absolute position of these non-phase-referenced observations. Fig.~\ref{WM1-4} shows that it is possible to shift the positions of these masers - applying an offset, $\Delta$RA = 25~mas and $\Delta$DEC = $-15$~mas, smaller than the positional uncertainty - such that they align with clusters 1~and~4. Since the excitation and the motion of the two maser clusters depend on very local conditions, this good agreement of their relative positions over observing epochs $\approx$20~yr apart confirms that the maser emission has been continuously excited at the clusters' loci for a long time. The water masers of clusters~1~and~4 have sky-projected velocities of $\approx$~10~\kms, which corresponds to a traveling distance of $\approx$~16~mas over 20~yr at the target distance of 2.7~kpc. If the maser velocities traced the effective speed of the shock fronts, we would have easily observed a significant relative displacement between the two clusters over the observing epochs (see Fig.~\ref{WM1-4}). However the masers move slowly enough to remain, during their lifetimes, sufficiently close to the places of persistent maser excitation to allow us to identify them.


 Our findings that the two extended shock fronts traced by clusters~1~and~4 are stationary can be naturally explained if the ionized gas hits a much denser layer of molecular material. Indeed, following the model by \citet{Mos14}, cluster~1 is adjacent to the inner portion (size $\approx$~200~au) of the disk surrounding the high-mass YSO IRS1a (see Figs.~\ref{MG}~and~\ref{G_stru}). 
 Although less straightforward, the interpretation of cluster~4 could be similar to the one for cluster~1, since cluster~4 is also found close to the disk midplane of another YSO, IRS1b (see Figs.~\ref{MG}~and~\ref{G_stru}). Following the model by \citet{Mos14}, the disk around IRS1b is thick and flaring out (opening angle of $\approx$~45\degree), which can explain the small offset of cluster~4 from the IRS1b's disk major axis on the sky. By studying the \Vlsr\ pattern of NH$_3$ absorption lines of different upper-state energies (400--1700~K), \citet{Mos14} have shown that the YSOs IRS1a and IRS1b (with their disks) are embedded inside the warm and dense envelope of molecular gas that surrounds the core of the HC~\HII\ region. The strongest absorption in all the NH$_3$ lines and the peak of dust continuum emission \citep{Beu13} correspond to the position of IRS1a, which suggests that this YSO could be the most massive one and the main one responsible for the ionization of the HC~\HII\ region.

Figures~\ref{MBG}~and~\ref{1_LS} show that cluster~1 marks the edge of the southern tip of the HC~\HII\ region's core, which protrudes toward the YSO IRS1a. Considering the expected direction of the outflow from IRS1a, we note that the position, orientation, and shape of this southern tip is consistent with that of an ionized cavity around this outflow. That is to say, an ionized cavity excavated by the IRS1a's outflow might naturally explain the position of the southern tip with respect to IRS1a. Models of high-mass star formation predict that the ejection of magnetocentrifugal, accretion-powered disk winds can carve bipolar cavities through the infalling envelope \citep{Sta23,Koe18,Kui16}, and these cavities would be photoionized for YSO masses of 10--20~\ms\ \citep{Tan16}. These cavities, with a typical density contrast of 100--1000, have parabolic shapes at scales of a few 100~au and conical shapes at scales of a few 1000~au owing to flow recollimation. Their opening angle stably increases with the YSO evolution, reaching values $>$ 60\degree\ for YSO masses $>$~20~\ms. Based on these models, we speculate that the water masers in cluster~1 trace the edge of the innermost portion of the outflow cavity near IRS1a and arise in shocks at the interface between the disk wind and the infalling envelope. In the following, we show that this interpretation is consistent with the water maser excitation models and accounts for the proper motions of the water masers in cluster~1.

The formation of the 22~GHz water masers in J-shocks requires a pre-shock number density in the range of \ 10$^6$--10$^8$~cm$^{-3}$ and a shock velocity (with respect to the pre-shock gas) of \ $\ge$30~\kms \ \citep{Hol13}. We note that the required shock velocity is significantly higher than the typical expansion speed of $\approx$~10~\kms\ of the ionized gas, which rules out the simple scenario in which the observed water masers are all excited by the expansion of the HC~\HII\ region. The J-shock model predicts that the thickness of the masing region (parallel to the shock velocity) essentially depends only on the pre-shock density, decreasing stably from 7 to 0.2~au with the number density increasing from 10$^6$ to 10$^8$~cm$^{-3}$ \citep[][see their Fig.~7]{Hol13}. From the image of the cluster~1 emission (see Fig.~\ref{G_stru}), we have calculated that the average maser thickness is 0.5~au, which is consistent with a pre-shock number density of\ $\gtrsim$~10$^7$~cm$^{-3}$. This value agrees with number densities of  \ $\sim$~10$^7$~cm$^{-3}$ \ predicted at the edge of the wind cavities at length scales of $\le$~100~au for YSOs with masses in the range of \ 15--30~\ms\ \citep[][see their Fig.~10 and Fig.~21a, respectively]{Tan16,Koe18}.

The water masers of cluster~1 could originate in shocks that arise when the wind impinges on the higher-density infalling material. In the rest-frame of the infalling gas, assuming momentum conservation, the shock speed is given by \ $V$~=~$\sqrt{\rho_{\rm win}/\rho_{\rm inf}}$~$\Upsilon$ \ \citep[see, for instance,][]{Mas93}. In this equation, \ $\rho_{\rm win}$ \ and \ $\Upsilon$ \ are the wind density and speed, respectively, and \ $\rho_{\rm inf}$ \ is the density of the infalling gas.
The Keplerian velocities provide a lower limit for the  velocities of both the magnetocentrifugal disk wind and the infalling material. Over the maser cluster~1, at radii $\le$100~au, we derive Keplerian velocities $\ge$~15~\kms\  taking for IRS1a the mass of 25~\ms\ estimated by \citet{Mos14}. Thus, the infall-relative velocity of the wind $\Upsilon$ is found to be $\ge$~30~\kms, as is required for the production of J-shocks.  Since the wind density can be a factor of $\sim$~100 lower than that of the infalling gas, the shock velocity $V$ (in the frame of the infalling gas) is inferred to be \ $\gtrsim$~3~\kms, which effectively means that the wind shocks against the dense infalling material and the shocked gas infalls at a speed reduced by a factor of $\approx$~(3/15)~$\approx$~ 20\%  with respect to that of the infalling gas. As we expect infall speeds of $\ge$~15~\kms,  the estimated shock velocities are consistent with the water maser proper motions in cluster~1, which have amplitudes of $\approx$~10~\kms\ and are approximately directed toward the YSO (see Fig.\ref{MBG}).

The shock front traced by the maser cluster~1 could also be produced by an outflow from IRS1b or an undetected YSO nearby IRS1a rather than by the wind-infall interaction discussed so far. We prefer the latter scenario for two main reasons, as follows. First, the regular spatial and 3D velocity distribution of the 6.7~GHz methanol masers around cluster~1 is well interpreted in terms of an edge-on rotating disk. We would not expect a linear distribution with aligned proper motions if what we observed was dense molecular material pushed by a fast outflow. A good example of the latter is provided by the HC~\HII\ region \G24~A1, where the ionized outflow ejected by an high-mass YSO impacts against surrounding dense molecular material \citep[][see their Figs.~9~and~10]{Mos18}. The water masers at the arc-like, terminal shock front of the ionized outflow move at speeds between 20~and~60~\kms, several times higher than what is observed in cluster~1, and the 6.7~GHz masers surround the shock front and trace the expansion of less shocked material at a speed of $\approx$~10~\kms.  Second, while the concomitant presence of a disk wind and infall at the position of IRS1a is readily explained, there would be no simple explanation for the positional correspondence of IRS1a with cluster~1 if the latter traced an outflow from a nearby YSO. On the other hand, cluster~4 is detached from the YSO IRS1b (see Fig.~\ref{G_stru}) and is also offset from the radio continuum; in this case, we cannot exclude that the shock front is generated by the impact of an outflow from a low-mass (undetected) YSO against the IRS1b's disk, rather than by the pressure of the ionized gas.

   \begin{figure*}
    \includegraphics[width=0.4\textwidth]{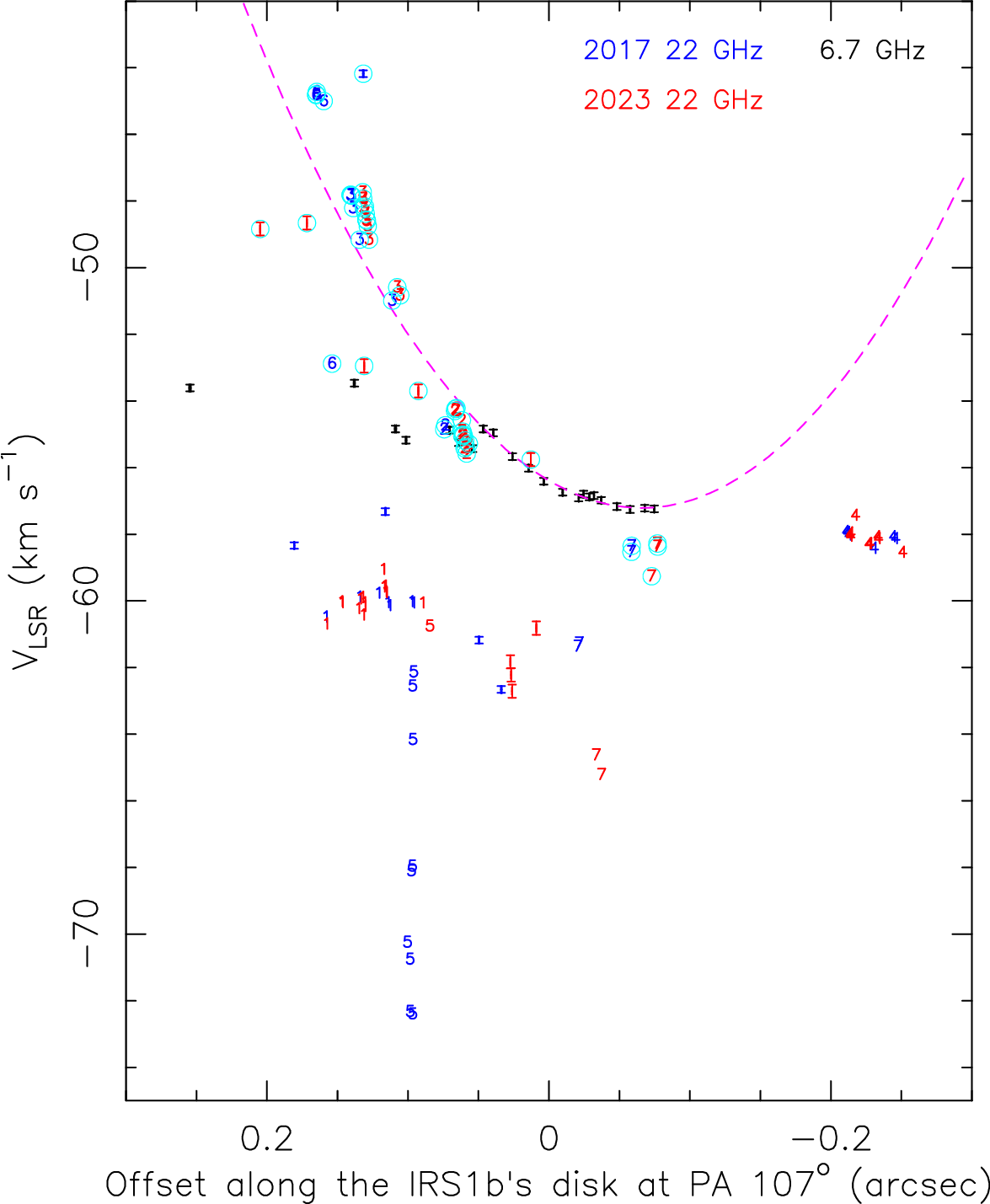} 
    \includegraphics[width=0.6\textwidth]{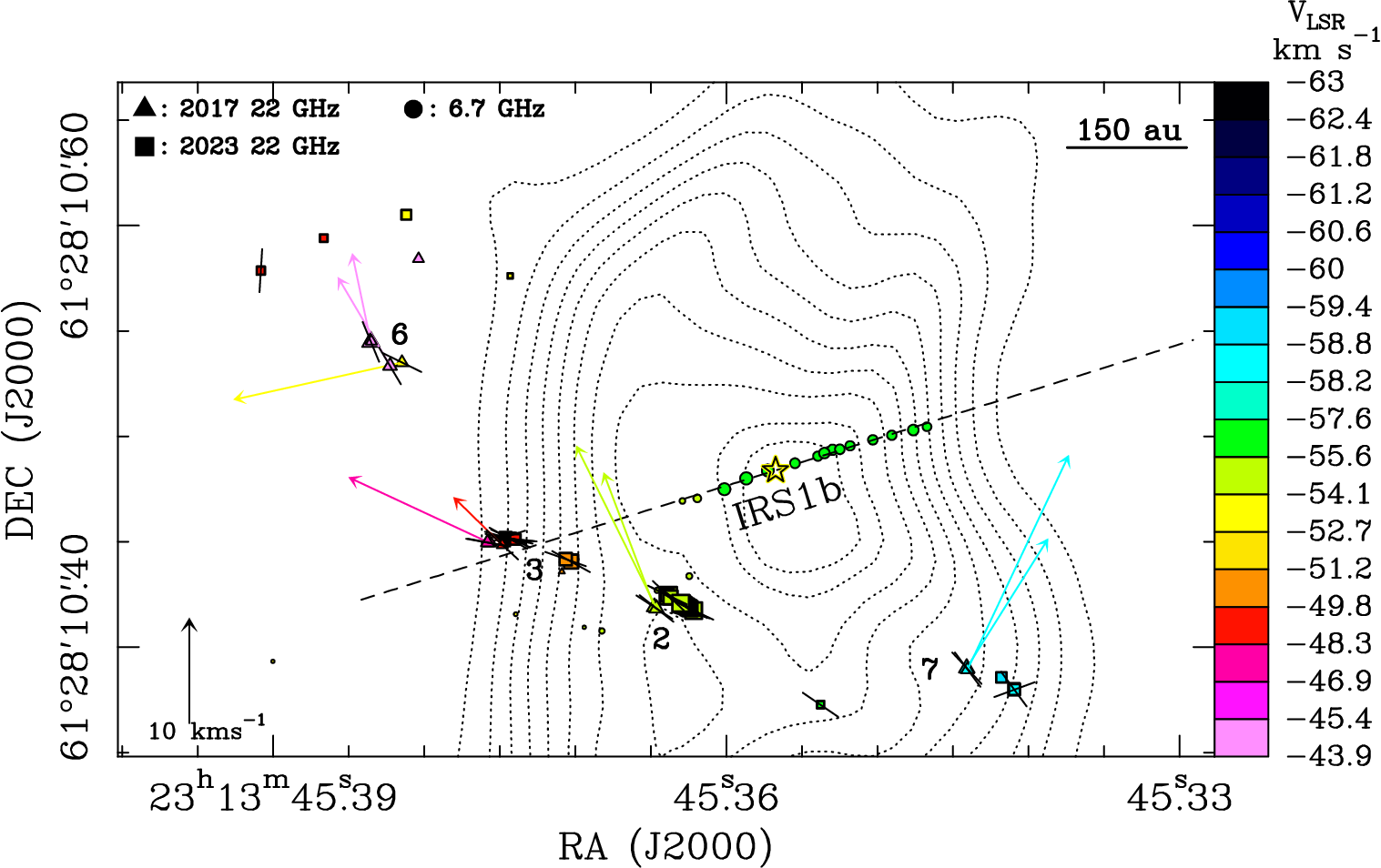}
      \caption{Water masers tracing the disk wind around IRS1b.\ (Left~panel)~Error bars and numbers give the \Vlsr\ of the 6.7~GHz methanol (black) and 22~GHz water (blue and red for the VLBA and global VLBI) masers plotted versus their position projected along the major axis of the IRS1b's disk at PA = 107\degree. The numbers are used to identify the water masers belonging to the clusters 1--7 (see Fig.~\ref{MBG}). Only the 6.7~GHz masers associated with IRS1b are shown. The magenta parabola gives the quadratic fit to the velocity profile of the 6.7~GHz masers from \citet{Mos14}, and the water masers that approximately follow the same velocity profile of the 6.7~GHz masers are encircled in cyan. \ (Right~panel)~Same as Fig.~\ref{MBG} but showing only the 22~GHz water masers encircled in cyan in the left~panel. The colored dots and the dashed line and black-yellow star have the same meaning as in Fig.~\ref{MG}.}
         \label{IRS1b}
   \end{figure*}

\subsection{The disk wind around IRS1b}

\label{DW}

\citet{Mos14} have reproduced the 3D velocities and the line-of-sight acceleration of the linear distribution of 6.7~GHz methanol masers in the northern core of the HC~\HII\ region with a model of edge-on disk in centrifugal equilibrium. A notable kinematic feature of the 6.7~GHz masers is that the change in their \Vlsr\ with position along the linear distribution is reproduced well with a quadratic curve. Figure~\ref{IRS1b}  (left~panel) shows that many water masers (from both the 2017 VLBA and 2023 global VLBI observations) also follow the same parabolic velocity profile of the 6.7~GHz masers. In this figure, we have encircled in cyan the water masers whose separation from the parabolic fit is either $\le$~0.1$^{\prime\prime}$ \  in position or $\le$~3~\kms\ in velocity. These water masers (as is evident by comparing the right panel of Fig.~\ref{IRS1b} with Fig.~\ref{MBG}) essentially correspond to all the ones detected near IRS1b, with the exception of the two clusters~4~and~5. In Sect.~\ref{inf}, we have already considered the case that cluster~4 is a stationary shock front against the IRS1b's disk. On the other hand, the proper motions of the water masers in cluster~5 point directly to the position of the YSO, suggesting that we might be observing a small clump of shocked molecular gas infalling through the ionized gas toward IRS1b.

The finding that most of the water masers detected close to IRS1b (i.e., the clusters 2, 3, 6, and 7) share the same \Vlsr\ pattern of the 6.7~GHz masers further probes the existence of a kinematic structure where both maser types originate. Following the model of \citet{Mos14} for the 6.7~GHz masers, we postulate that the water masers are also excited in the disk rotating around IRS1b. However, an important difference is that while the 6.7~GHz masers have proper motions aligned with their linear structure and easily interpretable as rotation \citep[][see their Fig.2]{Mos14}, the proper motions of the 22~GHz water masers form large angles with the line of 6.7~GHz masers (see Fig.~\ref{IRS1b}, right~panel), being on average directed (PA $\approx$~23\degree) roughly perpendicular to that. Besides, while the water masers have sky-projected distances between 230~and~530~au from IRS1b, their 3D velocities have amplitudes in the range of 10--23~\kms. By comparison, the rotation velocity around a central mass of 30~\ms\ (i.e., the stellar mass corresponding to the luminosity of the HC~\HII\ region and a stringent upper limit for the IRS1b mass) at a radius of 230~au is just $\approx$~10~\kms. It is thus clear that both the direction of motion and the speed of the water masers are inconsistent with rotation around IRS1b.

The agreement of the \Vlsr\ pattern between the 22~GHz water masers and typical tracers of disk rotation around YSOs (hot-core molecules like CH$_3$CN or CH$_3$OH, or the 6.7~GHz masers themselves)
and the concomitant deviation of the water maser proper motions (both in direction and amplitude) from rotation is a characteristic kinematic signature of a disk wind. By mapping the 3D velocities of the water masers at scales of 10--100~au from luminous YSOs, the POETS survey has found several cases of such kinematic behavior in the 22~GHz masers, of which the sources G11.92$-$0.61 and G35.02$+$0.35 are probably the clearest examples \citep[][see their Figs.~1~and~2]{Mos19}. In the former source, the presence of a disk wind has been fully confirmed by recent high-angular-resolution (beam of $\approx$~30~mas) ALMA observations \citep{Bay24}. The 3D velocities of the water masers can be explained, if the disk is seen sufficiently close to edge-on. In this case, while the \Vlsr\ reflect mainly the disk rotation, the proper motions are dominated by the poloidal velocities of the wind. 
The regular distribution in position and \Vlsr\ of the water maser clusters 2, 3, 6, and 7 around IRS1b indicates the presence of a rotating molecular disk that has a radius of several 100~au and that is probably thick or not exactly edge-on. The association with ionized gas, the large separation of the water masers from the YSO, and their relatively small speeds compared with those expected ($\ge$~20~\kms) from a magnetocentrifugal disk wind lead us to think that the water masers could emerge in a photo-evaporated disk wind \citep{Hol94,Ale14}. At a sufficiently large distance from the YSO, the thermal energy of the heated gas exceeds the gravitational binding energy and the gas can be centrifugally launched. Thus, the outflowing gas keeps the rotation signature of the disk.  The length scale of the launching region of a photo-evaporated wind is given by  the gravitational radius, \ $r_g = G \, M_{\rm YSO} \, / \, {c_s}^2 $, with  $G$ the gravitational constant, $M_{\rm YSO}$ \ the YSO mass, and \ $c_s$ \ the sound speed of 
the ionized gas (about \ 8--10~\kms\ in typical conditions).
For a high-mass YSO with $M_{\rm YSO}$ in the range of 10--30~\ms, $r_g$ varies between 100 and 300~au. The separations of the water masers from IRS1b are comparable with or exceed these values of $r_g$, and this is consistent with the scenario of a photo-evaporated disk wind.

\subsection{The disk-jet system around IRS1c}

\label{IRS1c}

Figure~\ref{1_LS} shows the map of the 1.3~continuum emission of the whole HC~\HII\ region. The positional correspondence of the continuum peak with IRS1b is an independent confirmation of the presence of this embedded YSO. At larger scales, $\gtrsim$~1000~au, the sensitive 2015 JVLA observations resolve the extended southern spherical component of the 1992 VLA image in three distinct elements: a spur of emission to the east of the IRS1a disk and two elongated narrow emissions. The morphology of the spur, stretching from north to south, might be associated with the direction of the outflow from IRS1b (traced by the dotted cyan line in Fig.~\ref{1_LS}). The other two emissions (traced by the dotted white lines in Fig.~\ref{1_LS}), more to the south, are perpendicular to each other. As was described in Sect.~\ref{arch}, we have shifted the JVLA continuum image (whose absolute position is uncertain) by a small offset to align the southernmost emission with the 6.7~GHz maser cluster placed $\approx$0\pas5 to the south of the HC~\HII\ region's core, in agreement with the 1992 VLA image (see Fig.~\ref{MG}).
IRS1a is clearly located at the southern border of the ionized core, where most of the gas of the region is concentrated. The dust continuum \citep[][see their Fig.~1]{Beu13} and NH$_3$ absorption \citep[][see their Fig.~10]{Mos14} maps show that the gas density and temperature at the position of the southern spherical component are substantially reduced with respect to IRS1a. The UV radiation from IRS1a could escape to the south and ionize gas concentrations at a relatively large distance. Thus, we speculate that the spur and the two elongated emissions are clumps of denser gas whose surface is ionized by IRS1a: 1)~the spur morphology hints at a collimated outflow emerging from IRS1b; \ 2)~the two other emissions to the south could reveal the disk-jet system associated with the YSO IRS1c, which \citet{Mos14} hypothesized to explain the excitation of the nearby 6.7~GHz masers. The presence of this YSO inside the southern spherical component is further indicated by the positive spectral index of the radio emission  \citep{Cam84,Gau95}. In agreement with (1), we note that the water masers are tracing a disk wind around IRS1b (see Sect.~\ref{DW}), which could eventually collimate into a jet. In line with (2), the radio continuum and the (radiatively pumped) 6.7~GHz masers are not symmetrically distributed around the YSO (at the intersection of the two dotted white lines), but their emissions are skewed toward the side closer to IRS1a (see Fig.~\ref{1_LS}), as if the radiation coming from IRS1a led to a differential excitation of the (putative) disk. At the \targ\ distance, the size and shape of the southernmost emission (the putative ionized disk) compare well with those of many "proplyds" observed with the JVLA in the Orion Nebula Cluster \citep[][see their Fig.~2]{Bal23}. In a few of these sources, faint radio emission that extends perpendicular to the proplyds could unveil the photoionized jet material, as we have proposed for the (putative) jet from IRS1c.
Finally, it is notable that the southeast-northwest direction of the (putative) jet from IRS1c agrees with that of the most compact  (on scales of \ $\sim$~10$^{\prime\prime}$) molecular outflow observed toward \targ\ \citep{Kam89,Dav98,Qiu11}.

\section{Conclusions}
\label{conclu}

This work employs VLBI observations of 22~GHz water masers to explore the gas kinematics close to the YSOs embedded in the HC~\HII\ region \targ. We extend the observational strategy of the POETS survey, which has recently studied, through the water masers on scales of 10--100~au, luminous YSOs prior to their ionization phase, to the case of more evolved, photoionizing YSOs. We have used multi-epoch VLBA observations to measure the 3D velocities of the water masers and more sensitive, global VLBI observations to map weaker maser emission and study the maser time variability.

Two linear distributions of 6.7~GHz methanol masers thread the southern and northern cores of the HC~\HII\ region, which we had previously modeled in terms of edge-on disks in centrifugal equilibrium around the YSOs IRS1a and IRS1b, respectively. The water masers trace two extended ($\ge$~100~au) arched shock fronts: one of them, the most prominent, adjacent to the inner portion of the disk around IRS1a, the other close to the midplane of the IRS1b's disk. The persistence of these shock fronts at invariant relative positions over $\approx$~20~yr attests that they are stationary. The shock front close to IRS1a marks the edge of the southern tip of the ionized core. We speculate that the southern tip might correspond to the ionized cavity of the disk wind ejected by IRS1a and the water masers arise in shocks at the edge of the cavity where the wind and the infalling envelope interact.
 
The water masers found close to IRS1b follow the same \Vlsr\ pattern as the 6.7~GHz methanol masers rotating in the disk. However, the proper motions of the water masers are totally inconsistent with disk rotation, both because they are directed at large angles from the disk midplane and because their amplitudes are far too large. This dichotomy of the water masers between their \Vlsr\ pattern, in agreement with the rotation profile of typical disk tracers in high-mass YSOs, and the proper motions, has been observed in many objects by the POETS survey and is a clear signature of a disk wind. In fact, if the disk is sufficiently close to edge-on, the \Vlsr\ mainly traces the disk rotation, and the proper motions the poloidal velocity of the wind. The association with ionized gas, the large distance (between 230 and 530~au) of the water masers from IRS1b and their speeds of 10--23~\kms\ (smaller than expected for a magnetohydrodynamic disk wind) lead us to favor the interpretation in terms of a photo-evaporated disk wind.

We have used archival 2015 JVLA data to produce a very sensitive image of the 1.3~cm continuum emission from the HC~\HII\ region \targ. The positional correspondences of the 1.3~cm continuum peak with IRS1b and of the southern tip of the ionized core with IRS1a are an independent confirmation of the presence of these two embedded YSOs. South of the core of the HC~\HII\ region, the 1.3~cm continuum image shows two narrow elongated emissions, perpendicular to each other and intercepting at a position inside a cluster of 6.7~GHz methanol masers. We interpret these structures as the disk-jet system of another YSO, IRS1c (located within the  6.7~GHz maser cluster), which becomes visible in its ionized emission following the UV illumination by IRS1a.

This work complements the results of the POETS survey showing that VLBI of the 22~GHz water masers can also trace the gas kinematics close to ionizing YSOs embedded in HC~\HII\ regions. Inside the core of the HC~\HII\ region \targ, our water maser observations support the presence of the two embedded YSOs IRS1a and IRS1b, previously identified through  VLBI of the 6.7~GHz methanol masers, by revealing a large shock front near the IRS1a's disk and the disk wind emitted by IRS1b. These results underline the importance of achieving a sufficiently high 
($\le$~100~au) linear resolution to resolve the kinematics of nearby YSOs.

\begin{acknowledgements}
CG acknowledges financial support by the Italian Ministry of University and Research (MUR)– Project CUP F53D23001260001, funded by the European Union – NextGenerationEU. This work was partially supported by FAPESP (Funda\c{c}\~ao de Amparo \'a Pesquisa do Estado de S\~ao Paulo) under grant 2021/01183-8.
TH is supported by the MEXT/JSPS KAKENHI Grant Numbers 17K05398, 18H05222, and 20H05845.

\end{acknowledgements}

%
   \bibliographystyle{aa} 
   \bibliography{biblio.bib} 

\begin{thebibliography}{49}
\expandafter\ifx\csname natexlab\endcsname\relax\def\natexlab#1{#1}\fi

\bibitem[{{Akabane} \& {Kuno}(2005)}]{Aka05}
{Akabane}, K. \& {Kuno}, N. 2005, \aap, 431, 183

\bibitem[{{Alexander} {et~al.}(2014){Alexander}, {Pascucci}, {Andrews},
  {Armitage}, \& {Cieza}}]{Ale14}
{Alexander}, R., {Pascucci}, I., {Andrews}, S., {Armitage}, P., \& {Cieza}, L.
  2014, in Protostars and Planets VI, ed. H.~{Beuther}, R.~S. {Klessen}, C.~P.
  {Dullemond}, \& T.~{Henning}, 475

\bibitem[{{Ballering} {et~al.}(2023){Ballering}, {Cleeves}, {Haworth}, {Bally},
  {Eisner}, {Ginsburg}, {Boyden}, {Fang}, \& {Kim}}]{Bal23}
{Ballering}, N.~P., {Cleeves}, L.~I., {Haworth}, T.~J., {et~al.} 2023, \apj,
  954, 127

\bibitem[{{Bayandina} {et~al.}(2024){Bayandina}, {Moscadelli}, {Cesaroni},
  {Beltr\'an}, {Sanna}, \& {Goddi}}]{Bay24}
{Bayandina}, O., {Moscadelli}, L., {Cesaroni}, R., {et~al.} 2024, \aap,
  submitted

\bibitem[{{Beuther} {et~al.}(2013){Beuther}, {Linz}, \& {Henning}}]{Beu13}
{Beuther}, H., {Linz}, H., \& {Henning}, T. 2013, \aap, 558, A81

\bibitem[{{Beuther} {et~al.}(2017){Beuther}, {Linz}, {Henning}, {Feng}, \&
  {Teague}}]{Beu17}
{Beuther}, H., {Linz}, H., {Henning}, T., {Feng}, S., \& {Teague}, R. 2017,
  \aap, 605, A61

\bibitem[{{Campbell}(1984)}]{Cam84}
{Campbell}, B. 1984, \apjl, 282, L27

\bibitem[{{Davis} {et~al.}(1998){Davis}, {Moriarty-Schieven}, {Eisl{\"o}ffel},
  {Hoare}, \& {Ray}}]{Dav98}
{Davis}, C.~J., {Moriarty-Schieven}, G., {Eisl{\"o}ffel}, J., {Hoare}, M.~G.,
  \& {Ray}, T.~P. 1998, \aj, 115, 1118

\bibitem[{{De Buizer} \& {Minier}(2005)}]{DeBui05}
{De Buizer}, J.~M. \& {Minier}, V. 2005, \apjl, 628, L151

\bibitem[{{De Pree} {et~al.}(2020){De Pree}, {Wilner}, {Kristensen},
  {Galv{\'a}n-Madrid}, {Goss}, {Klessen}, {Mac Low}, {Peters}, {Robinson},
  {Sloman}, \& {Rao}}]{DeP20}
{De Pree}, C.~G., {Wilner}, D.~J., {Kristensen}, L.~E., {et~al.} 2020, \aj,
  160, 234

\bibitem[{{Gaume} {et~al.}(1995){Gaume}, {Goss}, {Dickel}, {Wilson}, \&
  {Johnston}}]{Gau95}
{Gaume}, R.~A., {Goss}, W.~M., {Dickel}, H.~R., {Wilson}, T.~L., \& {Johnston},
  K.~J. 1995, \apj, 438, 776

\bibitem[{{Goddi} {et~al.}(2011){Goddi}, {Moscadelli}, \& {Sanna}}]{God11a}
{Goddi}, C., {Moscadelli}, L., \& {Sanna}, A. 2011, \aap, 535, L8

\bibitem[{{Greisen}(2003)}]{Gre03}
{Greisen}, E.~W. 2003, in Astrophysics and Space Science Library, Vol. 285,
  Information Handling in Astronomy - Historical Vistas, ed. A.~{Heck}, 109

\bibitem[{{Hollenbach} {et~al.}(2013){Hollenbach}, {Elitzur}, \&
  {McKee}}]{Hol13}
{Hollenbach}, D., {Elitzur}, M., \& {McKee}, C.~F. 2013, \apj, 773, 70

\bibitem[{{Hollenbach} {et~al.}(1994){Hollenbach}, {Johnstone}, {Lizano}, \&
  {Shu}}]{Hol94}
{Hollenbach}, D., {Johnstone}, D., {Lizano}, S., \& {Shu}, F. 1994, \apj, 428,
  654

\bibitem[{{Hosokawa} {et~al.}(2010){Hosokawa}, {Yorke}, \& {Omukai}}]{Hos10}
{Hosokawa}, T., {Yorke}, H.~W., \& {Omukai}, K. 2010, \apj, 721, 478

\bibitem[{{Kameya} {et~al.}(1989){Kameya}, {Hasegawa}, {Hirano}, {Takakubo}, \&
  {Seki}}]{Kam89}
{Kameya}, O., {Hasegawa}, T.~I., {Hirano}, N., {Takakubo}, K., \& {Seki}, M.
  1989, \apj, 339, 222

\bibitem[{{Keto}(2003)}]{Ket03}
{Keto}, E. 2003, \apj, 599, 1196

\bibitem[{{Keto} \& {Klaassen}(2008)}]{Ket08b}
{Keto}, E. \& {Klaassen}, P. 2008, \apjl, 678, L109

\bibitem[{{Keto} {et~al.}(2008){Keto}, {Zhang}, \& {Kurtz}}]{Ket08}
{Keto}, E., {Zhang}, Q., \& {Kurtz}, S. 2008, \apj, 672, 423

\bibitem[{{K{\"o}lligan} \& {Kuiper}(2018)}]{Koe18}
{K{\"o}lligan}, A. \& {Kuiper}, R. 2018, \aap, 620, A182

\bibitem[{{Kuiper} {et~al.}(2016){Kuiper}, {Turner}, \& {Yorke}}]{Kui16}
{Kuiper}, R., {Turner}, N.~J., \& {Yorke}, H.~W. 2016, \apj, 832, 40

\bibitem[{{Masson} \& {Chernin}(1993)}]{Mas93}
{Masson}, C.~R. \& {Chernin}, L.~M. 1993, \apj, 414, 230

\bibitem[{{Moscadelli} {et~al.}(2021){Moscadelli}, {Beuther}, {Ahmadi},
  {Gieser}, {Massi}, {Cesaroni}, {S{\'a}nchez-Monge}, {Bacciotti},
  {Beltr{\'a}n}, {Csengeri}, {Galv{\'a}n-Madrid}, {Henning}, {Klaassen},
  {Kuiper}, {Leurini}, {Longmore}, {Maud}, {M{\"o}ller}, {Palau}, {Peters},
  {Pudritz}, {Sanna}, {Semenov}, {Urquhart}, {Winters}, \& {Zinnecker}}]{Mos21}
{Moscadelli}, L., {Beuther}, H., {Ahmadi}, A., {et~al.} 2021, \aap, 647, A114

\bibitem[{{Moscadelli} {et~al.}(2011){Moscadelli}, {Cesaroni}, {Rioja},
  {Dodson}, \& {Reid}}]{Mos11a}
{Moscadelli}, L., {Cesaroni}, R., {Rioja}, M.~J., {Dodson}, R., \& {Reid},
  M.~J. 2011, \aap, 526, A66+

\bibitem[{{Moscadelli} \& {Goddi}(2014)}]{Mos14}
{Moscadelli}, L. \& {Goddi}, C. 2014, \aap, 566, A150

\bibitem[{{Moscadelli} {et~al.}(2007){Moscadelli}, {Goddi}, {Cesaroni},
  {Beltr{\'a}n}, \& {Furuya}}]{Mos07}
{Moscadelli}, L., {Goddi}, C., {Cesaroni}, R., {Beltr{\'a}n}, M.~T., \&
  {Furuya}, R.~S. 2007, \aap, 472, 867

\bibitem[{{Moscadelli} {et~al.}(2024){Moscadelli}, {Oliva}, {Sanna}, {Surcis},
  \& {Bayandina}}]{Mos24}
{Moscadelli}, L., {Oliva}, A., {Sanna}, A., {Surcis}, G., \& {Bayandina}, O.
  2024, \aap, 690, A81

\bibitem[{{Moscadelli} {et~al.}(2009){Moscadelli}, {Reid}, {Menten},
  {Brunthaler}, {Zheng}, \& {Xu}}]{Mos09}
{Moscadelli}, L., {Reid}, M.~J., {Menten}, K.~M., {et~al.} 2009, \apj, 693, 406

\bibitem[{{Moscadelli} {et~al.}(2018){Moscadelli}, {Rivilla}, {Cesaroni},
  {Beltr{\'a}n}, {S{\'a}nchez-Monge}, {Schilke}, {Mottram}, {Ahmadi}, {Allen},
  {Beuther}, {Csengeri}, {Etoka}, {Galli}, {Goddi}, {Johnston}, {Klaassen},
  {Kuiper}, {Kumar}, {Maud}, {M{\"o}ller}, {Peters}, {Van der Tak}, \&
  {Vig}}]{Mos18}
{Moscadelli}, L., {Rivilla}, V.~M., {Cesaroni}, R., {et~al.} 2018, \aap, 616,
  A66

\bibitem[{{Moscadelli} {et~al.}(2016){Moscadelli}, {S{\'a}nchez-Monge},
  {Goddi}, {Li}, {Sanna}, {Cesaroni}, {Pestalozzi}, {Molinari}, \&
  {Reid}}]{Mos16}
{Moscadelli}, L., {S{\'a}nchez-Monge}, {\'A}., {Goddi}, C., {et~al.} 2016,
  \aap, 585, A71

\bibitem[{{Moscadelli} {et~al.}(2019){Moscadelli}, {Sanna}, {Cesaroni},
  {Rivilla}, {Goddi}, \& {Rygl}}]{Mos19}
{Moscadelli}, L., {Sanna}, A., {Cesaroni}, R., {et~al.} 2019, \aap, 622, A206

\bibitem[{{Moscadelli} {et~al.}(2020){Moscadelli}, {Sanna}, {Goddi},
  {Krishnan}, {Massi}, \& {Bacciotti}}]{Mos20}
{Moscadelli}, L., {Sanna}, A., {Goddi}, C., {et~al.} 2020, \aap, 635, A118

\bibitem[{{Moscadelli} {et~al.}(2006){Moscadelli}, {Testi}, {Furuya}, {Goddi},
  {Claussen}, {Kitamura}, \& {Wootten}}]{Mos06}
{Moscadelli}, L., {Testi}, L., {Furuya}, R.~S., {et~al.} 2006, \aap, 446, 985

\bibitem[{{Pickett} {et~al.}(1998){Pickett}, {Poynter}, {Cohen}, {Delitsky},
  {Pearson}, \& {M{\"u}ller}}]{Pick98}
{Pickett}, H.~M., {Poynter}, R.~L., {Cohen}, E.~A., {et~al.} 1998, \jqsrt, 60,
  883

\bibitem[{{Qiu} {et~al.}(2011){Qiu}, {Zhang}, \& {Menten}}]{Qiu11}
{Qiu}, K., {Zhang}, Q., \& {Menten}, K.~M. 2011, \apj, 728, 6

\bibitem[{{Reid} {et~al.}(2009){Reid}, {Menten}, {Brunthaler}, {Zheng},
  {Moscadelli}, \& {Xu}}]{Rei09}
{Reid}, M.~J., {Menten}, K.~M., {Brunthaler}, A., {et~al.} 2009, \apj, 693, 397

\bibitem[{{Rivera-Soto} {et~al.}(2020){Rivera-Soto}, {Galv{\'a}n-Madrid},
  {Ginsburg}, \& {Kurtz}}]{Riv20}
{Rivera-Soto}, R., {Galv{\'a}n-Madrid}, R., {Ginsburg}, A., \& {Kurtz}, S.
  2020, \apj, 899, 94

\bibitem[{{Sandell} {et~al.}(2020){Sandell}, {Wright}, {G{\"u}sten},
  {Wiesemeyer}, {Reyes}, {Mookerjea}, \& {Corder}}]{Sand20}
{Sandell}, G., {Wright}, M., {G{\"u}sten}, R., {et~al.} 2020, \apj, 904, 139

\bibitem[{{Sanna} {et~al.}(2010){Sanna}, {Moscadelli}, {Cesaroni}, {Tarchi},
  {Furuya}, \& {Goddi}}]{San10b}
{Sanna}, A., {Moscadelli}, L., {Cesaroni}, R., {et~al.} 2010, \aap, 517, A78+

\bibitem[{{Sanna} {et~al.}(2018){Sanna}, {Moscadelli}, {Goddi}, {Krishnan}, \&
  {Massi}}]{San18}
{Sanna}, A., {Moscadelli}, L., {Goddi}, C., {Krishnan}, V., \& {Massi}, F.
  2018, \aap, 619, A107

\bibitem[{{Sewi{\l}o} {et~al.}(2008){Sewi{\l}o}, {Churchwell}, {Kurtz}, {Goss},
  \& {Hofner}}]{Sew08}
{Sewi{\l}o}, M., {Churchwell}, E., {Kurtz}, S., {Goss}, W.~M., \& {Hofner}, P.
  2008, \apj, 681, 350

\bibitem[{{Staff} {et~al.}(2023){Staff}, {Tanaka}, {Ramsey}, {Zhang}, \&
  {Tan}}]{Sta23}
{Staff}, J.~E., {Tanaka}, K. E.~I., {Ramsey}, J.~P., {Zhang}, Y., \& {Tan},
  J.~C. 2023, \apj, 947, 40

\bibitem[{{Surcis} {et~al.}(2011){Surcis}, {Vlemmings}, {Torres}, {van
  Langevelde}, \& {Hutawarakorn Kramer}}]{Sur11b}
{Surcis}, G., {Vlemmings}, W.~H.~T., {Torres}, R.~M., {van Langevelde}, H.~J.,
  \& {Hutawarakorn Kramer}, B. 2011, \aap, 533, A47

\bibitem[{{Tan} {et~al.}(2014){Tan}, {Beltr{\'a}n}, {Caselli}, {Fontani},
  {Fuente}, {Krumholz}, {McKee}, \& {Stolte}}]{Tan14}
{Tan}, J.~C., {Beltr{\'a}n}, M.~T., {Caselli}, P., {et~al.} 2014, Protostars
  and Planets VI, 149

\bibitem[{{Tanaka} {et~al.}(2016){Tanaka}, {Tan}, \& {Zhang}}]{Tan16}
{Tanaka}, K. E.~I., {Tan}, J.~C., \& {Zhang}, Y. 2016, \apj, 818, 52

\bibitem[{{Wood} \& {Churchwell}(1989)}]{Woo89}
{Wood}, D. O.~S. \& {Churchwell}, E. 1989, \apj, 340, 265

\bibitem[{{Zhang} {et~al.}(2019){Zhang}, {Tanaka}, {Rosero}, {Tan}, {Marvil},
  {Cheng}, {Liu}, {Beltr{\'a}n}, \& {Garay}}]{ZhaY19}
{Zhang}, Y., {Tanaka}, K. E.~I., {Rosero}, V., {et~al.} 2019, \apjl, 886, L4

\bibitem[{{Zhu} {et~al.}(2013){Zhu}, {Zhao}, {Wright}, {Sandell}, {Shi}, {Wu},
  {Brogan}, \& {Corder}}]{Zhu13}
{Zhu}, L., {Zhao}, J.-H., {Wright}, M.~C.~H., {et~al.} 2013, \apj, 779, 51

\end{thebibliography}

\begin{appendix}
\onecolumn

\section{Parameters of the H$_2$O masers}

The following two tables list the parameters of the H$_2$O masers from the VLBA and global VLBI observations.

\begin{longtable}{ccccrrrr} 
\caption{\label{wat1} 22.2~GHz H$_2$O maser parameters for \targ\ from the VLBA observations.}\\  
\hline\hline
Feature & Epochs\tablefootmark{a} of & I$_{\rm peak}$ & $V_{\rm LSR}$ & \multicolumn{1}{c}{$\Delta~x$} & \multicolumn{1}{c}{$\Delta~y$} & \multicolumn{1}{c}{$V_{x}$} & \multicolumn{1}{c}{$V_{y}$} \\
Number  & Detection & (Jy beam$^{-1}$) & (km s$^{-1}$) & \multicolumn{1}{c}{(mas)} & \multicolumn{1}{c}{(mas)} & \multicolumn{1}{c}{(km s$^{-1}$)} & \multicolumn{1}{c}{(km s$^{-1}$)} \\
\hline
\endfirsthead
\caption{continued.}\\
\hline\hline
Feature & Epochs\tablefootmark{a} of & I$_{\rm peak}$ & $V_{\rm LSR}$ & \multicolumn{1}{c}{$\Delta~x$} & \multicolumn{1}{c}{$\Delta~y$} & \multicolumn{1}{c}{$V_{x}$} & \multicolumn{1}{c}{$V_{y}$} \\
Number  & Detection & (Jy beam$^{-1}$) & (km s$^{-1}$) & \multicolumn{1}{c}{(mas)} & \multicolumn{1}{c}{(mas)} & \multicolumn{1}{c}{(km s$^{-1}$)} & \multicolumn{1}{c}{(km s$^{-1}$)} \\
\hline
\endhead
\hline
\endfoot
\hline
\endlastfoot
      1 &        1,2,3,4 &    34.365 &   $-$59.8 &     0.00\hspace*{0.4cm} &     0.00\hspace*{0.4cm} &   $-$6.7$\pm$ 7.7 &   $-$4.5$\pm$ 3.3  \\
    2 &        1,2,3,4 &     8.297 &   $-$68.1 &    58.68$\pm$0.18 &   265.25$\pm$0.09 &  $-$17.5$\pm$ 3.2 &   $-$2.6$\pm$ 3.7  \\
    3 &        1,2,3,4 &     5.114 &   $-$44.8 &   126.55$\pm$0.18 &   256.04$\pm$0.08 &    1.8$\pm$ 3.1 &    8.4$\pm$ 3.5  \\
    4 &        1,2,3,4 &     2.181 &   $-$60.0 &   $-$23.48$\pm$0.19 &     3.85$\pm$0.08 &    8.7$\pm$ 5.5 &   $-$4.4$\pm$ 3.3  \\
    5 &        1,2,3,4 &     1.975 &   $-$58.5 &  $-$155.99$\pm$0.19 &   101.29$\pm$0.10 &   $-$7.7$\pm$ 3.6 &   12.3$\pm$ 3.8  \\
    6 &        1,2,3,4 &     1.774 &   $-$59.9 &    14.38$\pm$0.21 &     0.75$\pm$0.07 &   11.6$\pm$ 5.7 &   $-$3.4$\pm$ 3.0  \\
    7 &              1 &     1.522 &   $-$57.3 &    85.87$\pm$0.18 &   290.30$\pm$0.08 &    ....\hspace*{0.5cm}  &    ....\hspace*{0.5cm}   \\
    8 &        1,2,3,4 &     1.434 &   $-$58.0 &  $-$273.93$\pm$0.19 &   238.08$\pm$0.08 &   10.3$\pm$ 3.8 &   $-$3.1$\pm$ 3.4  \\
    9 &        1,2,3,4 &     1.336 &   $-$54.8 &    $-$7.84$\pm$0.23 &   130.22$\pm$0.16 &    4.8$\pm$ 7.1 &   12.8$\pm$ 7.0  \\
   10 &        1,2,3,4 &     1.329 &   $-$58.3 &    78.18$\pm$0.18 &    48.68$\pm$0.08 &    5.1$\pm$ 3.2 &   $-$7.1$\pm$ 3.2  \\
   11 &        1,2,3,4 &     1.220 &   $-$58.4 &  $-$294.56$\pm$0.19 &   234.78$\pm$0.08 &   $-$2.6$\pm$ 3.8 &   $-$9.5$\pm$ 3.0  \\
   12 &        1,2,3,4 &     1.132 &   $-$54.7 &    $-$8.87$\pm$0.23 &   129.59$\pm$0.17 &    7.6$\pm$ 6.6 &   15.5$\pm$ 6.5  \\
   13 &        1,2,3,4 &     0.832 &   $-$61.4 &  $-$116.87$\pm$0.20 &    99.38$\pm$0.10 &   $-$4.8$\pm$ 4.8 &   $-$1.6$\pm$ 4.3  \\
   14 &        1,2,3,4 &     0.726 &   $-$45.0 &   117.47$\pm$0.18 &   244.67$\pm$0.08 &    5.0$\pm$ 3.3 &    8.4$\pm$ 3.3  \\
   15 &              4 &     0.629 &   $-$68.0 &    57.87$\pm$0.16 &   265.25$\pm$0.10 &    ....\hspace*{0.5cm}  &    ....\hspace*{0.5cm}   \\
   16 &              2 &     0.573 &   $-$47.8 &    71.21$\pm$0.16 &   160.97$\pm$0.07 &    ....\hspace*{0.5cm}  &    ....\hspace*{0.5cm}   \\
   17 &        1,2,3,4 &     0.567 &   $-$58.0 &  $-$275.15$\pm$0.18 &   237.21$\pm$0.07 &   18.5$\pm$ 3.2 &    0.8$\pm$ 3.1  \\
   18 &          1,3,4 &     0.560 &   $-$47.8 &    70.44$\pm$0.20 &   160.74$\pm$0.08 &   13.3$\pm$ 4.4 &    6.2$\pm$ 3.3  \\
   19 &        1,2,3,4 &     0.553 &   $-$58.1 &  $-$308.01$\pm$0.18 &   242.56$\pm$0.09 &   $-$8.2$\pm$ 3.4 &   $-$0.7$\pm$ 3.8  \\
   20 &        1,2,3,4 &     0.504 &   $-$58.3 &  $-$156.39$\pm$0.19 &   100.59$\pm$0.21 &   $-$9.6$\pm$ 3.8 &   20.4$\pm$ 7.2  \\
   21 &          1,2,3 &     0.478 &   $-$60.1 &    $-$6.33$\pm$0.18 &     1.69$\pm$0.08 &   $-$2.8$\pm$ 5.6 &   $-$3.5$\pm$ 4.6  \\
   22 &          1,2,3 &     0.436 &   $-$52.9 &   111.75$\pm$0.18 &   246.21$\pm$0.07 &   16.0$\pm$ 4.4 &   $-$3.5$\pm$ 4.3  \\
   23 &            1,3 &     0.424 &   $-$60.1 &   $-$24.60$\pm$0.19 &     4.16$\pm$0.08 &    ....\hspace*{0.5cm}  &    ....\hspace*{0.5cm}   \\
   24 &        1,2,3,4 &     0.398 &   $-$60.5 &    42.14$\pm$0.20 &     9.34$\pm$0.10 &    1.7$\pm$ 5.2 &   $-$7.7$\pm$ 4.2  \\
   25 &            1,3 &     0.369 &   $-$47.8 &    71.28$\pm$0.24 &   160.82$\pm$0.08 &    ....\hspace*{0.5cm}  &    ....\hspace*{0.5cm}   \\
   26 &        1,2,3,4 &     0.348 &   $-$60.2 &    $-$7.58$\pm$0.24 &     2.09$\pm$0.10 &   $-$7.8$\pm$ 7.9 &   $-$2.0$\pm$ 4.4  \\
   27 &        1,2,3,4 &     0.313 &   $-$70.2 &    59.65$\pm$0.20 &   259.02$\pm$0.21 &  $-$13.0$\pm$ 4.3 &  $-$10.4$\pm$ 6.7  \\
   28 &        1,2,3,4 &     0.273 &   $-$58.1 &  $-$306.73$\pm$0.19 &   241.71$\pm$0.08 &   $-$0.8$\pm$ 4.2 &   $-$9.4$\pm$ 3.8  \\
   29 &        1,2,3,4 &     0.246 &   $-$64.2 &    57.82$\pm$0.18 &   266.09$\pm$0.08 &   $-$2.7$\pm$ 4.0 &    0.5$\pm$ 3.5  \\
   30 &        1,2,3,4 &     0.244 &   $-$49.2 &    64.24$\pm$0.19 &   159.73$\pm$0.08 &    4.6$\pm$ 3.6 &    4.5$\pm$ 3.2  \\
   31 &              4 &     0.240 &   $-$44.7 &   126.32$\pm$0.16 &   256.34$\pm$0.13 &    ....\hspace*{0.5cm}  &    ....\hspace*{0.5cm}   \\
   32 &              1 &     0.239 &   $-$57.9 &  $-$272.70$\pm$0.18 &   238.90$\pm$0.08 &    ....\hspace*{0.5cm}  &    ....\hspace*{0.5cm}   \\
   33 &              1 &     0.233 &   $-$44.2 &   103.95$\pm$0.18 &   295.47$\pm$0.08 &    ....\hspace*{0.5cm}  &    ....\hspace*{0.5cm}   \\
   34 &              3 &     0.202 &   $-$44.8 &   126.87$\pm$0.17 &   257.15$\pm$0.08 &    ....\hspace*{0.5cm}  &    ....\hspace*{0.5cm}   \\
   35 &            2,3 &     0.182 &   $-$72.4 &    57.78$\pm$0.15 &   266.88$\pm$0.08 &    ....\hspace*{0.5cm}  &    ....\hspace*{0.5cm}   \\
   36 &              1 &     0.156 &   $-$72.3 &    58.88$\pm$0.19 &   263.14$\pm$0.09 &    ....\hspace*{0.5cm}  &    ....\hspace*{0.5cm}   \\
   37 &              1 &     0.130 &   $-$62.6 &    58.04$\pm$0.19 &   266.20$\pm$0.09 &    ....\hspace*{0.5cm}  &    ....\hspace*{0.5cm}   \\
   38 &              1 &     0.119 &   $-$48.2 &    69.26$\pm$0.18 &   160.61$\pm$0.08 &    ....\hspace*{0.5cm}  &    ....\hspace*{0.5cm}   \\
   39 &              2 &     0.100 &   $-$62.1 &    57.33$\pm$0.17 &   266.63$\pm$0.09 &    ....\hspace*{0.5cm}  &    ....\hspace*{0.5cm}   \\
   40 &        1,2,3,4 &     0.099 &   $-$61.3 &  $-$117.63$\pm$0.20 &    99.95$\pm$0.08 &   $-$7.0$\pm$ 4.7 &   $-$1.5$\pm$ 3.4  \\
   41 &              1 &     0.097 &   $-$62.7 &   $-$90.12$\pm$0.19 &     2.59$\pm$0.09 &    ....\hspace*{0.5cm}  &    ....\hspace*{0.5cm}   \\
   42 &              1 &     0.067 &   $-$70.8 &    58.46$\pm$0.19 &   261.42$\pm$0.11 &    ....\hspace*{0.5cm}  &    ....\hspace*{0.5cm}   \\
   43 &            2,3 &     0.059 &   $-$51.0 &    36.51$\pm$0.20 &   147.29$\pm$0.12 &    ....\hspace*{0.5cm}  &    ....\hspace*{0.5cm}   \\
   44 &              2 &     0.046 &   $-$61.2 &   $-$73.79$\pm$0.24 &     1.84$\pm$0.14 &    ....\hspace*{0.5cm}  &    ....\hspace*{0.5cm}   \\

\end{longtable} 
\tablefoot{
\\
\tablefoottext{a}{The VLBA epochs are: 1)~March 2, 2017; \ 2)~May 27, 2017; \ 3)~June 11, 2017; \ 4)~July 22, 2017.} \\
Column~1 gives the feature label number; Column~2 lists the observing epochs at which the feature was detected;
Columns~3~and~4 provide the intensity of the strongest spot
and the intensity-weighted \Vlsr, respectively, averaged over the
observing epochs; Columns~5~and~6 give the position offsets (with
the associated errors) along the RA and DEC axes, relative to feature~\#1, measured at the first epoch of detection; Columns~7~and~8 give the components of the absolute proper motion (with the associated errors) along the RA and DEC axes.\\
The absolute position of the feature~\#1 at the first observing epoch on March 2, 2017, is: 
RA~(J2000) = 23$^{\rm h}$ 13$^{\rm m}$ 45\fs36587, DEC~(J2000) = 61\degree 28$^{\prime}$ 10\farcs2888, with an accuracy of \ $\pm$0.5~mas. 
}   

\begin{longtable}{cccrr} 
\caption{\label{wat2} 22.2~GHz H$_2$O maser parameters for \targ\ from the global VLBI observations.}\\  
\hline\hline
Feature &  I$_{\rm peak}$ & $V_{\rm LSR}$ & \multicolumn{1}{c}{$\Delta~x$} & \multicolumn{1}{c}{$\Delta~y$} \\
Number  &  (Jy beam$^{-1}$) & (km s$^{-1}$) & \multicolumn{1}{c}{(mas)} & \multicolumn{1}{c}{(mas)}  \\
\hline
\endfirsthead
\caption{continued.}\\
\hline\hline
Feature &  I$_{\rm peak}$ & $V_{\rm LSR}$ & \multicolumn{1}{c}{$\Delta~x$} & \multicolumn{1}{c}{$\Delta~y$} \\
Number  & (Jy beam$^{-1}$) & (km s$^{-1}$) & \multicolumn{1}{c}{(mas)} & \multicolumn{1}{c}{(mas)}  \\
\hline
\endhead
\hline
\endfoot
\hline
\endlastfoot
    1 &    28.345 &   $-$59.7 &     0.00\hspace*{0.4cm} &     0.00\hspace*{0.4cm}  \\
    2 &    19.009 &   $-$59.6 &     0.45$\pm$0.09 &    $-$0.06$\pm$0.04  \\
    3 &    11.678 &   $-$59.8 &    $-$0.36$\pm$0.09 &     0.06$\pm$0.04  \\
    4 &     3.588 &   $-$58.1 &  $-$271.34$\pm$0.08 &   235.80$\pm$0.07  \\
    5 &     3.364 &   $-$58.1 &  $-$293.49$\pm$0.10 &   231.52$\pm$0.04  \\
    6 &     2.846 &   $-$54.2 &    $-$8.97$\pm$0.09 &   136.39$\pm$0.07  \\
    7 &     2.786 &   $-$60.0 &    19.07$\pm$0.07 &     0.54$\pm$0.04  \\
    8 &     2.115 &   $-$59.6 &     1.21$\pm$0.07 &    $-$0.17$\pm$0.04  \\
    9 &     2.010 &   $-$60.1 &   $-$25.74$\pm$0.08 &     5.46$\pm$0.04  \\
   10 &     1.463 &   $-$59.9 &    17.31$\pm$0.12 &     0.17$\pm$0.04  \\
   11 &     1.252 &   $-$60.1 &    33.68$\pm$0.11 &     4.34$\pm$0.04  \\
   12 &     1.102 &   $-$60.0 &    34.14$\pm$0.15 &     4.51$\pm$0.08  \\
   13 &     1.035 &   $-$54.9 &   $-$15.45$\pm$0.15 &   132.39$\pm$0.10  \\
   14 &     1.032 &   $-$58.1 &  $-$293.07$\pm$0.14 &   231.36$\pm$0.05  \\
   15 &     1.020 &   $-$55.0 &   $-$15.11$\pm$0.09 &   132.59$\pm$0.06  \\
   16 &     0.952 &   $-$54.6 &   $-$14.81$\pm$0.10 &   132.84$\pm$0.06  \\
   17 &     0.748 &   $-$64.6 &  $-$118.91$\pm$0.07 &   119.46$\pm$0.04  \\
   18 &     0.668 &   $-$55.0 &   $-$15.98$\pm$0.10 &   132.08$\pm$0.06  \\
   19 &     0.605 &   $-$60.1 &    15.75$\pm$0.08 &    $-$0.11$\pm$0.04  \\
   20 &     0.534 &   $-$55.0 &   $-$16.49$\pm$0.10 &   131.83$\pm$0.05  \\
   21 &     0.355 &   $-$60.2 &    15.43$\pm$0.10 &    $-$0.16$\pm$0.04  \\
   22 &     0.335 &   $-$61.8 &   $-$86.35$\pm$0.09 &    19.38$\pm$0.06  \\
   23 &     0.282 &   $-$62.2 &   $-$86.77$\pm$0.08 &    19.70$\pm$0.05  \\
   24 &     0.244 &   $-$55.2 &   $-$17.80$\pm$0.10 &   131.18$\pm$0.05  \\
   25 &     0.235 &   $-$55.4 &   $-$17.37$\pm$0.07 &   131.39$\pm$0.04  \\
   26 &     0.213 &   $-$58.3 &  $-$287.78$\pm$0.07 &   229.84$\pm$0.04  \\
   27 &     0.211 &   $-$58.0 &  $-$271.80$\pm$0.12 &   235.44$\pm$0.05  \\
   28 &     0.164 &   $-$47.9 &    68.21$\pm$0.07 &   163.89$\pm$0.03  \\
   29 &     0.148 &   $-$50.8 &    37.19$\pm$0.08 &   152.60$\pm$0.06  \\
   30 &     0.147 &   $-$60.2 &    20.43$\pm$0.07 &     0.83$\pm$0.04  \\
   31 &     0.146 &   $-$58.3 &  $-$172.85$\pm$0.07 &    92.14$\pm$0.04  \\
   32 &     0.144 &   $-$54.3 &    $-$8.54$\pm$0.08 &   136.70$\pm$0.04  \\
   33 &     0.142 &   $-$55.4 &   $-$18.46$\pm$0.08 &   130.73$\pm$0.04  \\
   34 &     0.138 &   $-$58.3 &  $-$287.02$\pm$0.07 &   229.75$\pm$0.04  \\
   35 &     0.134 &   $-$58.4 &  $-$173.12$\pm$0.07 &    91.83$\pm$0.04  \\
   36 &     0.093 &   $-$58.0 &  $-$270.92$\pm$0.08 &   235.85$\pm$0.11  \\
   37 &     0.087 &   $-$54.2 &    $-$9.97$\pm$0.10 &   135.59$\pm$0.08  \\
   38 &     0.074 &   $-$53.0 &   115.30$\pm$0.07 &   317.09$\pm$0.04  \\
   39 &     0.070 &   $-$48.6 &    65.77$\pm$0.07 &   163.51$\pm$0.04  \\
   40 &     0.062 &   $-$59.3 &  $-$166.85$\pm$0.08 &    97.62$\pm$0.06  \\
   41 &     0.056 &   $-$48.4 &    66.29$\pm$0.08 &   163.65$\pm$0.04  \\
   42 &     0.054 &   $-$59.1 &     1.68$\pm$0.07 &    $-$0.22$\pm$0.04  \\
   43 &     0.053 &   $-$55.3 &   $-$21.06$\pm$0.11 &   129.37$\pm$0.05  \\
   44 &     0.053 &   $-$60.4 &    16.75$\pm$0.07 &     0.11$\pm$0.04  \\
   45 &     0.052 &   $-$48.8 &    64.62$\pm$0.09 &   163.33$\pm$0.04  \\
   46 &     0.048 &   $-$55.6 &   $-$19.20$\pm$0.23 &   130.27$\pm$0.11  \\
   47 &     0.048 &   $-$48.2 &    67.66$\pm$0.07 &   163.75$\pm$0.04  \\
   48 &     0.039 &   $-$48.6 &    65.30$\pm$0.07 &   163.37$\pm$0.04  \\
   49 &     0.037 &   $-$60.7 &    47.19$\pm$0.13 &    10.17$\pm$0.09  \\
   50 &     0.036 &   $-$48.8 &   184.27$\pm$0.07 &   290.62$\pm$0.04  \\
   51 &     0.035 &   $-$49.2 &    63.64$\pm$0.09 &   163.11$\pm$0.04  \\
   52 &     0.030 &   $-$48.7 &   154.41$\pm$0.07 &   306.04$\pm$0.04  \\
   53 &     0.024 &   $-$62.7 &   $-$87.32$\pm$0.08 &    20.19$\pm$0.04  \\
   54 &     0.023 &   $-$60.8 &    47.78$\pm$0.07 &   257.47$\pm$0.04  \\
   55 &     0.023 &   $-$58.5 &  $-$308.17$\pm$0.09 &   242.06$\pm$0.07  \\
   56 &     0.015 &   $-$65.2 &  $-$122.22$\pm$0.07 &   121.42$\pm$0.04  \\
   57 &     0.015 &   $-$53.7 &    66.01$\pm$0.07 &   288.18$\pm$0.04  \\
   58 &     0.014 &   $-$47.7 &    68.59$\pm$0.07 &   164.05$\pm$0.04  \\
   59 &     0.013 &   $-$57.4 &  $-$273.91$\pm$0.07 &   239.51$\pm$0.04  \\
   60 &     0.012 &   $-$50.6 &    39.74$\pm$0.13 &   153.76$\pm$0.06  \\
   61 &     0.011 &   $-$48.2 &    66.87$\pm$0.08 &   163.70$\pm$0.04  \\
   62 &     0.011 &   $-$60.8 &   $-$89.37$\pm$0.07 &    71.59$\pm$0.05  \\
   63 &     0.010 &   $-$55.8 &   $-$81.13$\pm$0.08 &    84.69$\pm$0.05  \\

\end{longtable} 
\tablefoot{
\\
Column~1 gives the feature label number; 
Columns~2~and~3 provide the intensity of the strongest spot
and the intensity-weighted \Vlsr, respectively; Columns~4~and~5 give the position offsets (with
the associated errors) along the RA and DEC axes, relative to feature~\#1.\\
The absolute position of the feature~\#1 at the observing epoch on June 5, 2023, is: 
RA~(J2000) = 23$^{\rm h}$ 13$^{\rm m}$ 45\fs36302, DEC~(J2000) = 61\degree 28$^{\prime}$ 10\farcs2730, with an accuracy of \ $\pm$0.5~mas. 
}   

\end{appendix}

\end{document}